\documentclass{aa}

\usepackage{color}
\usepackage{graphicx}
\usepackage{txfonts}
\usepackage{longtable}
\usepackage{lscape}

\begin{document}

   \title{Chemical fingerprints of hot Jupiter planet formation
    \thanks{
    Based on data products from observations made with ESO Telescopes at the La Silla Paranal Observatory
    under programme ID 072.C-0033(A), 072.C-0488(E), 074.B-0455(A), 075.C-0202(A), 077.C-0192(A),           
    077.D-0525(A), 078.C-0378(A), 078.C-0378(B), 080.A-9021(A), 082.C-0312(A)
    082.C-0446(A), 083.A-9003(A), 083.A-9011(A), 083.A-9011(B), 083.A-9013(A), 083.C-0794(A), 084.A-9003(A),          
    084.A-9004(B), 085.A-9027(A), 085.C-0743(A),  087.A-9008(A),  088.C-0892(A), 089.C-0440(A), 089.C-0444(A),           
    089.C-0732(A), 090.C-0345(A), 092.A-9002(A), 192.C-0852(A), 60.A-9036(A), 60.A-9120(B), and
    60.A-9700(A); and on data products from the SHOPIE archive.           
    }
    }   
%    \fnmsep
%         \thanks{
%             Tables~\ref{parameters_table_full} and ~\ref{abundance_table_full}
%                  are only available in electronic
%                       format.
%                            }
%                            }     

   \subtitle{ } 

   \author{J. Maldonado
          \inst{1}
          \and
          E. Villaver  
          \inst{2}
          \and
          C. Eiroa 
          \inst{2}
          }

          \institute{INAF - Osservatorio Astronomico di Palermo,
                     Piazza del Parlamento 1, 90134 Palermo, Italy
          \and
           Universidad Aut\'onoma de Madrid, Dpto. F\'isica Te\'orica, M\'odulo 15,
                         Facultad de Ciencias, Campus de Cantoblanco, 28049 Madrid, Spain
  %        \and
  %         stop mi hada estrella invitada
            }   

   \offprints{J. Maldonado \\ \email{jmaldonado@astropa.inaf.it}}
   \date{Received September 15, 1996; accepted March 16, 1997}

% \abstract{}{}{}{}{} 
% 5 {} token are mandatory
 
  \abstract
   {
   The current paradigm to explain the presence of Jupiter-like planets with small orbital periods
   (P $<$ 10 days; hot Jupiters), which involves their formation beyond the snow line following
   inward migration, has been challenged by recent works that explore the possibility 
   of {\it in situ} formation.
   }
   {
   We aim to test whether stars harbouring hot Jupiters and stars with more distant
   gas-giant planets show any chemical peculiarity that could be related to different formation processes.
   }  
   {
    Our methodology is based on the analysis of high-resolution \'echelle spectra.
    Stellar parameters and abundances of C, O, Na, Mg, Al, Si, S, Ca, Sc, Ti, V, Cr, Mn, Co, Ni, Cu, and Zn
    for a  sample of 88 planet hosts are derived. The sample is divided into stars hosting hot (a $<$ 0.1 au) and
    cool  (a $>$ 0.1 au) Jupiter-like planets.
    The metallicity and abundance trends of the two sub-samples are compared and set in the context of current models
    of planet formation and migration.
   }
  {
  Our results show that stars with hot Jupiters have higher metallicities than stars with cool
  distant gas-giant planets in the metallicity range  +0.00/+0.20 dex. The data also shows a tendency
  of stars with cool Jupiters to show larger abundances of $\alpha$ elements. % and volatile elements. 
  No abundance differences between stars with cool and hot Jupiters are found when considering 
  iron peak,  volatile elements or the C/O, and Mg/Si ratios.
   The corresponding $p$-values from the statistical tests comparing
  the cumulative distributions of cool and hot planet hosts are
  0.20, $<$ 0.01, 0.81, and 0.16 for metallicity, $\alpha$, iron-peak,
  and volatile elements, respectively.
  We confirm previous works suggesting that
  more distant planets show higher planetary masses as well as larger eccentricities.
   We note differences in age and spectral type between the  hot and cool planet host samples that might 
  affect the abundance comparison.
 }
  {
   The differences in the distribution of planetary mass, period, eccentricity, and stellar host
   metallicity suggest a different formation mechanism for hot and cool Jupiters.
  % In particular, an {\it in situ} formation at short orbital distances would explain the lower
  % volatile abundances found in stars with hot Jupiters {\bf although this hypothesis needs further confirmation.} 
   The slightly larger $\alpha$ abundances found in stars harbouring cool Jupiters might compensate their
   lower metallicities allowing the formation of gas-giant planets. 
  }


   \maketitle
%
%<<<<<<<<<<<<<<<<<<<<<<<<<<<<<<<<<<<<<<<<<<<<<<<<<<<<<<<<<<<<<<<<
\section{Introduction}
\label{introduccion}
%>>>>>>>>>>>>>>>>>>>>>>>>>>>>>>>>>>>>>>>>>>>>>>>>>>>>>>>>>>>>>>>>

 The discovery of gas-giant planets orbiting their parent stars at close distances, the so-called hot Jupiters 
 \citep[e.g.][]{1995Natur.378..355M,1997ApJ...474L.115B},
 was an unexpected surprise, as  no planets like these
 are present in our own solar system.
 It was noticed early on that the gas temperature in the inner
 region of the protoplanetary disc was too high to allow
 for the condensation of solid particles and also that the available mass so close
 to the star was not enough to form a Jupiter mass object
 \citep{1995Sci...267..360B,2000Icar..143....2B,2006ApJ...648..666R}. 
 Therefore, it was proposed that all Jupiter-size planets form
  at relatively low temperatures beyond the snowline
% at large
% distances from the star (3-5 au) 
 and that hot Jupiters experience subsequent
 inward migration to their current locations \citep{1996Natur.380..606L}.
 Currently, migration mechanisms are divided into two main categories:
 models that involve disc migration %% and require the truncation of the disc to explain the period distribution
 \citep[e.g.][]{1980ApJ...241..425G,1996Icar..124...62P,1997Icar..126..261W,2000MNRAS.315..823P,2005A&A...434..343A,
  2008ApJ...673..487I,2009A&A...501.1139M,2011ApJ...735...29B,2012ARA&A..50..211K},
  and models that involve secular high planet eccentricity that need a mechanism to drive it,
  involving either planet-planet encounters 
 \citep[e.g.][]{1996Sci...274..954R,2006ApJ...638L..45F,2008ApJ...686..621F,
  2008ApJ...686..580C,2008ApJ...686..603J,2010ApJ...725.1995M,2011ApJ...742...72N,
      2012ApJ...751..119B,2012ApJ...754...57B}, secular Lidov-Kozai (LK) oscillations
      \citep{1962P&SS....9..719L,1962AJ.....67R.579K}
   induced by a companion  \citep[e.g.][]{2003ApJ...589..605W,2007ApJ...669.1298F,2012ApJ...754L..36N},
    or  secular chaos \citep{2011ApJ...735..109W,2016MNRAS.455.3180H}.

  The formation of gas-giant planets at wide stellar separations by core-accretion
 faces the obstacle of an excessively long timescale \citep[e.g.][]{2009ApJ...707...79D,2011ApJ...727...86R}.
 The so-called pebble accretion mechanism (in which small particles with sizes ranging from millimetres to centimetres
 constitute the main drivers of planetary growth)
 has been shown to accelerate the growth speed of planetary cores
 \citep{2010MNRAS.404..475J,2010A&A...520A..43O,2012A&A...544A..32L,2014A&A...572A.107L,2017AREPS..45..359J}.
% As the core growths it carves a gap on the disc which interrupts the flux of pebbles on to the core
% \citep{2014A&A...572A..35L}.
% This happens when the core reaches a mass of approximately
% 10 M$_{\oplus}$ and the contraction of the gas envelope initiates.
% During the growth process, gas planets migrate over distances comparable to their initial semimajor axis. % while they grow
% \citep[see][and references therein]{2017AREPS..45..359J}.
 Population synthesis models based on pebble accretion strongly suggest that snowlines are the
 preferred places for planet formation \citep{2017MNRAS.469.5016A}. %Pebble accretion models
% also predict a significant diversity in the planetary
% C/O, O/H ratios according to different formation and migration mechanisms  \citep{2017MNRAS.469.4102M}.

 However, recently, several works have addressed the question of whether
 the {\it in situ} formation through core accretion of hot Jupiters might be possible \citep{2016ApJ...829..114B,2016ApJ...817L..17B}.
 The reason is twofold. Firstly, the lack of enough material in the innermost region of the protoplanetary nebula
 is mainly based on models of the solar nebula 
 and it is possible that the solar nebula is not representative of the general conditions of disc forming planets
 \citep[e.g.][]{2013MNRAS.431.3444C}.
 %{\bf although it should be noted that the protoplanetary disc would always
 %consist of a hot inner region and a coulder retion.} 
 In particular, small planets  (mini-Neptunes or Super-Earths) not present in the solar system,
 have been found to be quite common  in multiple systems at small orbital periods
 \citep{2011arXiv1109.2497M,2012Natur.481..167C,2012ApJS..201...15H,2013ApJ...766...81F,2013PNAS..11019273P}.
 Although stating the obvious, it is important to recall that irrespective of whether or not the innermost
 parts of the disks are suitable for planet formation it is expected that in the colder parts of the disk, solid condensation should be important.

 Secondly,
 evidence has accumulated over recent years that hot Jupiters and more distant gas-giant planets have different properties. 
 To start with, their frequencies are different.
 It has been shown that even after accounting for possible observational  biases
 \citep{2014ApJ...785..126K,2016ApJ...821...89B}, there are two distinct peaks in the orbital
 period distribution of Jupiter-class planets: one for hot Jupiters (representing 0.5-1.0\% of Sun-type
 stars \citep{2012ApJ...753..160W})  at  orbital  periods  close  to  three  days  and
 the average giant planet population that  shows periods of 100 days $<$ $P$ $<$ 3000 days
 (representing 10.5\% of the nearby FGK stars \citep{2008PASP..120..531C}). 
 Between these distances the planetary distribution shows a clear dip or ``valley''
 \citep[e.g.][]{2003MNRAS.341..948J,2009ApJ...693.1084W,2009ApJ...694L.171C}. %%%,2003MNRAS.341..948J}.
  In a recent work, \cite{2016A&A...587A..64S} analysed a sample of 129 giant planets with periods less than 400 days
 taken from the {\sc Kepler} catalogue of transit candidates.
 The authors find a frequency of hot Jupiters around FGK stars in the {\sc Kepler} data in the range 
 0.4-0.5\% while the occurrence rate from radial velocity surveys is of the order of $\sim$ 1\%.
 Such a difference might be related to the gas-giant planet-metallicity correlation
 as the {\sc Kepler} sample is found to differ in metallicity by about 0.15-0.20 dex from the solar
 neighbourhood. A similar discrepancy is also found for giant planets in the period valley,
 with frequencies of  0.90\% \citep{2016A&A...587A..64S}, and  $\sim$ 1.6\%
 \citep{2013ApJ...766...81F,2011arXiv1109.2497M}.
 For planets with periods longer than $\sim$ 85 days, the occurrence estimates
 from the different surveys are compatible
 with an occurrence rate of $\sim$ 3\%.

 Hot Jupiters show masses below that of, or of the order of,   Jupiter. On the other hand, cool distant gas-giant planets 
 have masses of several times Jupiter's mass \citep[e.g.,][]{2016ApJ...829..114B}.
 Furthermore, a tendency for the eccentricities to increase with
 the planet mass has also been noticed
 \citep[see][and references therein]{2007A&A...464..779R}.
 The occurrence of additional long-period planets in stars with hot Jupiters 
 is still debated. 
 \cite{2016ApJ...825...62S} found no difference in the companion fraction
 between hot and cool Jupiters, either inside or outside the water-ice line, 
 a result that is inconsistent with the simplest models of high-eccentricity 
 migration. Concurrently, \cite{2017arXiv170607807D} using data from the {\sc Kepler}
 mission show that hot Jupiters as well as hot Neptunes 
 are preferentially found in systems with one transiting planet,
 in agreement with previous works also based on {\sc Kepler} data 
 \citep{2011ApJ...732L..24L,2012PNAS..109.7982S,2016ApJ...825...98H}.   
 These results, however, should be regarded with caution as it has been argued that transit
 detection might be biased towards short-period planets \citep[e.g.][]{2016ApJ...825...62S}.

 Regarding the metallicity abundance of planet hosts, although no clear correlations
 between the stellar metallicity and the planetary semimajor axis were found
 \citep{2007ARA&A..45..397U,2008PhST..130a4003V,2009A&A...507..523A},
 several authors have noted a tendency of stars with hot Jupiters planets to show
 slightly larger metallicities than stars with more distant planets
 \citep{2004MNRAS.354.1194S}, especially in the
 ``low''-metallicity range (between -0.50 dex and + 0.00 dex; \citep{2015A&A...579A..20M}). 
 Furthermore, \cite{2015A&A...579A..20M} note that there are no hot Jupiters harbouring stars with significant low metallicities (below -0.50/-0.60 dex),
 while some cool Jupiters can be found around stars as metal-poor as -1.00 dex.
 Several works have suggested that the architecture of a planet 
 (period and eccentricity) might indeed show a dependence on the
 host star's metallicity \citep[][]{2007A&A...464..779R,2013ApJ...767L..24D,2013ApJ...763...12B,2013A&A...560A..51A}.
 More recently, \cite{2017arXiv170107654B} and \cite{2017A&A...603A..30S} suggest that stars hosting massive gas-giant planets 
 show on average lower metallicities than the stars hosting planets
 with masses below $\sim$ 4-5 M$_{\rm Jup}$. 
  A possible correlation between planetary mass and stellar metallicity have been 
 also suggested by \cite{2017MNRAS.466..443J}.

 We believe that our current knowledge on hot-Jupiter formation mechanisms
 would benefit from a detailed chemical abundance analysis of a large sample of stars
 harbouring hot and cool gas-giant planets, including a large number of ions besides iron.
 Therefore, in this paper we present a homogeneous analysis of a large sample of stars hosting gas-giant planets that
 is based on high-resolution and high signal-to-noise ratio \'echelle spectra. 

 This paper is organised as follows. Section~\ref{section_observations}
 describes the stellar sample, the observations, and the spectroscopic
 analysis. The comparison of the abundances of stars with hot and cool Jupiters
 is presented in Sect.~\ref{seccion_analysis} where
 a search for correlations between the stellar abundances and the planetary
 properties is also presented.
 The results are set in the context of current planet formation models
 in Sect.~\ref{seccion_discussion}.
 Our conclusions follow in Sect.~\ref{conclusions}.

%<<<<<<<<<<<<<<<<<<<<<<<<<<<<<<<<<<<<<<<<<<<<<<<<<<<<<<<<<<<<<<<<<<
\section{Data and spectroscopic analysis}
\label{section_observations}
%>>>>>>>>>>>>>>>>>>>>>>>>>>>>>>>>>>>>>>>>>>>>>>>>>>>>>>>>>>>>>>>>>>
\subsection{Stellar sample and observations}
% -----------------------------------------------------------------

 Two samples of main sequence (MS) stars, one of cool gas-giant
 planet hosts (hereafter cool PHs, semimajor axis a $>$ 0.1 au), and another one of stars harbouring 
 hot gas-giant planets (hot PHs, semimajor axis a $<$ 0.1 au), were built by carefully checking
 the data available\footnote{Up to February 2017}
 at the Extrasolar Planets Encyclopaedia\footnote{http://exoplanet.eu/}
 \citep{2011A&A...532A..79S}. 
 % We note that
  Only radial velocity planets were considered for this work % as metallicity differences
  % between stars targeted in radial velocity and transit surveys have been
  % reported.
  as transit targets are known to have systematically lower metallicities than
  radial velocity targets \citep{2006AcA....56....1G,2012ApJS..201...15H,2012ApJ...753..160W,2013ApJ...767L..24D}
  and there is no bias that prevents RV detection at small orbital periods.
  %%% . references is ....pJS..201...15H
 The division at a = 0.1 au comes from the known distribution of planetary semimajor
 axis where an enhanced frequency of close-in gas-giant planets
 is found at semimajor axes $\leq$ 0.07 au \citep[e.g.][]{2009ApJ...693.1084W,2009ApJ...694L.171C}.
 Stars with multiple planets are classified as hot PHs if at least
 one of the planets has a semimajor axis lower than 0.1 au.
 We note, however, that the pileup of planets at $\sim$ 3 days 
 is present when the full data set of known planets
 (dominated by transit surveys) is considered, while it is not
 obvious when only radial velocity or {\sc Kepler} data is
 analysed \citep{2016ApJ...817L..17B}.

 For this work, all planetary hosts with spectral types between F5 and K2
 and available values of planetary mass (M$_{\rm p}$$\sin i$), and semimajor axis were
 considered. Stars with low-mass planets  (M$_{\rm p}$$\sin i$ $<$ 30M$_{\oplus}$)
 are not included since these stars
 have been the focus of exhaustive analysis and do not
 present metallicity signatures similar to the ones found for stars hosting gas-giant planets
 \citep{2010ApJ...720.1290G,2011arXiv1109.2497M,2011A&A...533A.141S,2012Natur.486..375B,2015ApJ...808..187B}.
 Objects with masses (M$_{\rm p}$$\sin i$ $>$ 13M$_{\rm Jup}$)
 have also not been included in this analysis since their masses enter into the brown dwarf domain,
 their formation mechanism might be different altogether from planets, and the analysis of their 
 abundances have been the subject of a separate study \citep{2017A&A...602A..38M}.
 Whether or not giant stars with planets follow the gas-giant planet metallicity
 correlation seen in MS planet hosts has been the subject of recent discussions
 \citep{
 2005PASJ...57..127S,2005ApJ...632L.131S,2007A&A...475.1003H,2007A&A...473..979P,2008PASJ...60..781T,2010ApJ...725..721G,
 2013A&A...554A..84M,2013A&A...557A..70M,2015A&A...574A..50J,2015A&A...574A.116R,2016A&A...588A..98M}.
 We therefore excluded all stars with $\log g$ values (see following Section) lower than 4.0 (cgs).

 The high-resolution spectra used in this work were collected from public archives.
 FEROS \citep{1999Msngr..95....8K}  data from the ESO archive\footnote{http://archive.eso.org/wdb/wdb/adp/phase3\_main/form}
 were taken for 74 stars, while for 8 stars HARPS spectra were used
 \citep{2003Msngr.114...20M}. 
 Additional data for 5 stars was taken from the SOPHIE \citep{2006tafp.conf..319B} archive
 \footnote{http://atlas.obs-hp.fr/sophie/} while
 for the star HIP 7513 ($\upsilon$ And) a  spectrum taken with 
 the 2dcoud\'e spectrograph at the McDonald Observatory \citep{1995PASP..107..251T}
 from the public library ``S$^{\rm 4}$N'' \citep{2004A&A...420..183A} 
 was analysed. 
 All spectra were reduced by the corresponding pipelines which implement the typical
 corrections involved in \'echelle spectra reduction.
 The spectra were corrected for radial velocity shifts by using
 radial velocity standard stars or the radial velocities provided by
 the reduction pipelines. Typical values of the signal-to-noise ratio
 are between 80 and 230 (measured in the region around 605 nm).
 The properties of the different spectra are summarised in Table~\ref{tabla_espectrografos}.
 
%--------------------------------------------------------------------------------
%  --- Table 1: Properties of the different spectrographs
%-------------------------------------------------------------------------------
\begin{table}
\centering
\caption{
Properties of the different spectrographs used in this work.
}
\label{tabla_espectrografos}
\begin{scriptsize}
\begin{tabular}{lccc}
\hline\noalign{\smallskip}
Spectrograph &  Spectral range (\AA)  & Resolving power  & $N$ stars \\
\hline %\noaling{\smallskip}
 FEROS        &  3500-9200             &   48000        & 74  \\
 HARPS        &  3780-6910             &  115000        & 8   \\
 SOPHIE       &  3872-6943             &   75000        & 5   \\
 2dcoud\'e    &  3400-10000            &   60000        & 1   \\        
 \hline
 \end{tabular}
 \end{scriptsize}
 \end{table}

 The total number of stars in the cool-PHs amounts to 59:
 4 F-type stars, 44 G-type stars, and 11 K-type stars. 
 Hot-PHs account for 29 stars: 7 F-type stars, 13
 G-type stars, and 9 K-type stars. 
 The stars are listed in Table~\ref{parameters_table_full}.
  All the planetary data used in this work -- minimum mass, 
 semimajor axis, planetary period, and eccentricity
 (Sect.~\ref{seccion_analysis}) 
 -- come from
 the Extrasolar Planets Encyclopaedia 
 \citep{2011A&A...532A..79S}.

%+++++++++++++++++++++++++++++++++++++++++++++++++++++++++++++
% --- Table 2: Stellar properties and parameters
%+++++++++++++++++++++++++++++++++++++++++++++++++++++++++++++
%\input{tablas/hot_cool_stars_tabla_parameters}

%+++++++++++++++++++++++++++++++++++++++++++++++++++++++++++++++++++
\subsection{Spectroscopic analysis}
%+++++++++++++++++++++++++++++++++++++++++++++++++++++++++++++++++++
 
  Basic stellar parameters T$_{\rm eff}$, $\log g$, microturbulent
  velocity $\xi_{\rm t}$, and [Fe/H] are determined using the code
  {\sc TGVIT} %\footnote{http://optik2.mtk.nao.ac.jp/\textasciitilde{}takeda/tgv/}
  \citep{2005PASJ...57...27T}, which implements the iron ionisation %, equilibrium
  and excitation balance, a methodology
  widely applied to solar-like stars. % (write some examples here).
  The line list as well as the adopted parameters (excitation potential,
  $\log (gf)$ values) can be found on Y. Takeda's web page
  \footnote{http://optik2.mtk.nao.ac.jp/\textasciitilde{}takeda/tgv/}.
  This code
  makes use of ATLAS9, plane-parallel, LTE atmosphere models
  \citep{1993KurCD..13.....K}.

  TGVIT derives the uncertainties in the stellar parameters
  by progressively changing each stellar
  parameter from the converged solution to a value in which the excitation
  equilibrium, the match of the curve of grow, or the ionisation equilibrium
  condition are no longer fulfilled \citep{2002PASJ...54..451T}.

  Stellar evolutionary parameters, age, mass, and radius, were computed from
  {\sc Hipparcos} V magnitudes \citep{1997ESASP1200.....E} and
  parallaxes \citep{2007A&A...474..653V} using the code
  PARAM\footnote{http://stev.oapd.inaf.it/cgi-bin/param} \citep{2006A&A...458..609D},
  which is based on the use of Bayesian methods, together with
  the PARSEC set of isochrones \citep{2012MNRAS.427..127B}.
  These parameters are provided in Table~\ref{parameters_table_full}.
 
  Chemical abundance of individual elements
  C, O, Na, Mg, Al, Si, S, Ca, Sc, Ti, V, Cr,  Mn, Co, Ni, Cu, and Zn
  was obtained using the 2014 version of the
  code {\sc MOOG}\footnote{http://www.as.utexas.edu/\textasciitilde{}chris/moog.html}
  \citep{1973PhDT.......180S} together with ATLAS9 atmosphere models
  \citep{1993KurCD..13.....K}. The measured equivalent widths (EWs) of a list of narrow,
  non-blended lines for each of the aforementioned species are used as inputs. The selected
  lines are taken from the list provided by \cite{2015A&A...579A..20M}.
  Hyperfine structure (HFS) was taken into account for
  V~{\sc i}, Co~{\sc i}, and Cu~{\sc i} abundances.
  Oxygen abundances derived from O~{\sc i} triplet lines at 777 nm are known to be severely
  affected by departures from LTE \citep[e.g.][]{1993A&A...275..269K,2001NewAR..45..559K}.
  To account for non-LTE effects, the prescriptions given by
  \cite{2003A&A...402..343T} were followed.
 
  Our derived abundances are provided in Table~\ref{abundance_table_full}.
  They are expressed relative to the solar values derived in \cite{2015A&A...579A..20M}.
  Errors take into account the line-to-line scatter as well as
  the propagation of the uncertainties of the stellar parameters
  in the abundances following the results by \cite{2015A&A...579A..20M}. 

\begin{figure}[htb]
  \centering
  \includegraphics[angle=270,scale=0.45]{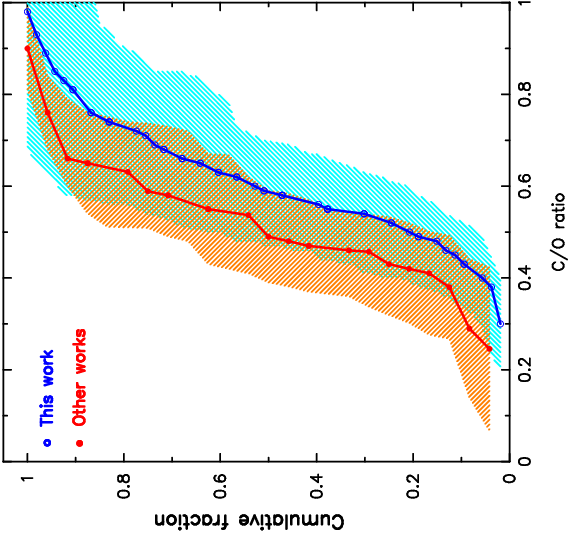}
  \caption{ 
  Comparison between the C/Os derived in this work and those
  from \cite{2014ApJ...788...39T} and \cite{2015ApJ...815....5S}.
  }
  \label{co_diagram}
  \end{figure}

   Given the difficulties of obtaining reliable abundances of some volatile
  elements, in particular C~{\sc i} and O~{\sc i} , we compared our derived C and O
  ratios (C/Os) with the works of \cite{2014ApJ...788...39T} and \cite{2015ApJ...815....5S}.
  Since a star-to-star comparison is not possible, as no stars in common are available,
  we compare the corresponding cumulative distribution function; Figure~\ref{co_diagram}.
  We restrict the comparison to stars with C/Os lower than one (to use the same
  range of values than the literature samples). We also discard for the analysis those stars 
  with high %%%%and discard stars with high
  uncertainties in the derived C/O values. The comparison reveals a tendency of 
  slightly larger values in our sample. A 
   Kolmogorv-Smirnov test
    (hereafter K-S test)\footnote{Performed with  the IDL Astronomy User's Library
     routine {\sc kstwo}, see  http://idlastro.gsfc.nasa.gov/} 
  returns a K-S statistics of 0.34, and a  $p$-value of 0.03 (n$_{\rm eff}$ = 16.5).

 We do not know the reason for this discrepancy. While it might be related to
 some difference in the methodology to derive abundances (e.g. line selection,
 atomic data), it could also be an effect
 of the different samples considered in this work.
 We note that the samples in \cite{2014ApJ...788...39T} and \cite{2015ApJ...815....5S}
 consist in stars with transiting planets, while this work is focused on radial
 velocity planets.

  %%------------------------------------------------------
  %  Figure 9: C/O ratios
  %%------------------------------------------------------

%++++++++++++++++++++++++++++++++++++++++++++++++++++++++++++++++++
% --- Tabla 3: Abundancias
%++++++++++++++++++++++++++++++++++++++++++++++++++++++++++++++++++
%\input{tablas/hotcool_table_abundancias}

%++++++++++++++++++++++++++++++++++++++++++++++++++++++++++++++++++
\subsection{Kinematics}
%++++++++++++++++++++++++++++++++++++++++++++++++++++++++++++++++++

 Galactic spatial velocity components $(U, V, W)$ are computed
 following the procedure described in \cite{2001MNRAS.328...45M}. %, 2010A&A...521A..12M
 Parallaxes were taken from the revised reduction of the 
 {\sc Hipparcos} data \citep{2007A&A...474..653V}  while proper motions are from the
 Tycho-2 catalogue \citep{2000A&A...357..367H}.
 Radial velocities are mainly from the compilation by \cite{2006yCat.3249....0M}.
 %or
 %derived by cross correlating the observed spectra with the spectra of radial velocity
 %standards.
 Finally, stars were classified as belonging to the thin/thick disk applying the methodology described in
 \cite{2003A&A...410..527B,2005A&A...433..185B}.
 This classification is provided in Table~\ref{parameters_table_full}.

% ----------------------------------------------------------------------
\subsection{Properties of the stellar samples}\label{biases}
% ----------------------------------------------------------------------

 It is well known that some stellar properties, such as distance, age, or kinematics, can
 affect the metal content of a star.
 Therefore before a comparison of metallicity and abundances of hot and cool PHs
 is made, an analysis of the stellar properties of both samples is mandatory. 
 This analysis is presented in Table~\ref{bias_table} where both samples are compared
 by using a  K-S test. 
% Kolmogorv-Smirnov test
% (hereafter K-S test)\footnote{Performed with  the IDL Astronomy User's Library
% routine {\sc kstwo}, see  http://idlastro.gsfc.nasa.gov/}.
  In order to account for the uncertainties in the stellar parameters
  a series of 10$^{\rm 6}$ simulations were performed. In each simulation the
  property of each star is randomly varied within its corresponding uncertainty.
  For each series of simulated data a K-S test is performed.
  Assuming that the distribution of the simulated $p$-values
  follows a Gaussian function we then compute the probability that
  the simulated $p$-value takes the value (within 5\%) found when analysing the original
  data.

\begin{figure}[!htb]
\centering
\begin{minipage}{0.49\linewidth}
\includegraphics[angle=270,scale=0.45]{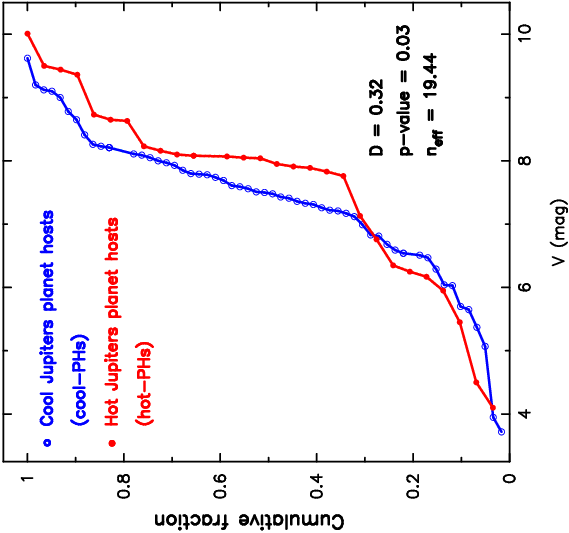}
%hc_stars_Vmag_distribution.ps}
\end{minipage}
\begin{minipage}{0.49\linewidth}
\includegraphics[angle=270,scale=0.45]{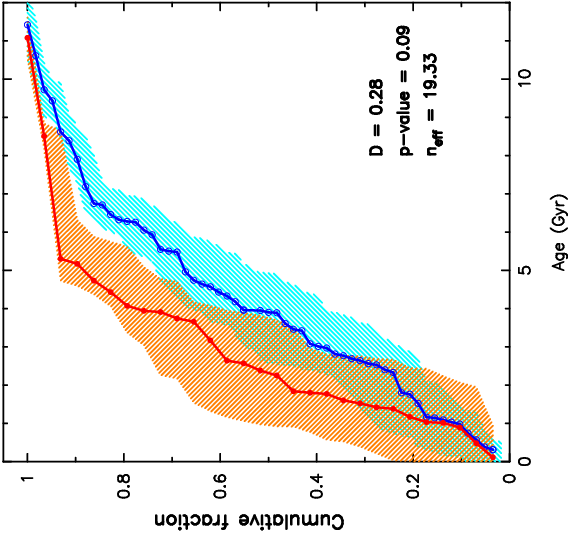}
%hc_stars_age_distribution.ps}
\end{minipage}
\begin{minipage}{0.49\linewidth}
\includegraphics[angle=270,scale=0.45]{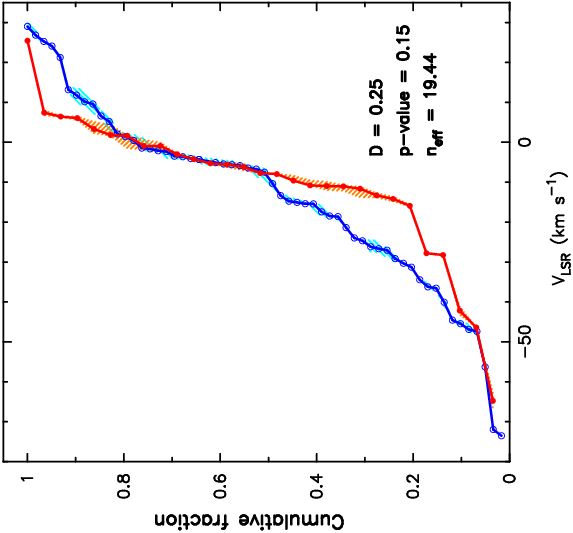}
%hc_stars_Vlsr_distribution.ps}
\end{minipage}
\begin{minipage}{0.49\linewidth}
\includegraphics[angle=270,scale=0.45]{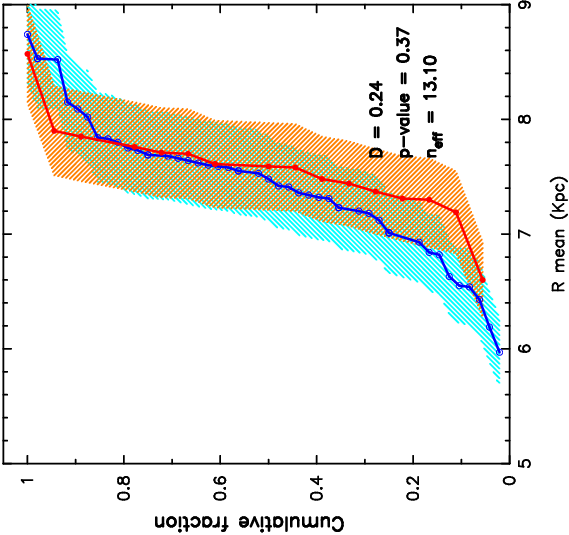}
%hc_stars_Rmean_distribution.ps}
\end{minipage}
\caption{ 
Cumulative distribution function of V mag (upper left panel), stellar ages (upper right panel),
$V_{\rm LSR}$ (bottom left), and mean Galactocentric distance (bottom right) for
cool and hot PHs.}
\label{cummu_bias}
\end{figure}

 According to the KS test, both samples are quite similar in terms of stellar distance. The analysis
 reveals a tendency of hot PHs to be slightly fainter and younger than
 the cool PHs. This effect can also be seen in Figure~\ref{cummu_bias}
 where the corresponding cumulative distribution functions of the V magnitude (upper left panel)
 and the stellar age (upper right panel) are shown.
  The hatched regions indicate the limits of the cumulative distributions
 built by considering that all stars have a stellar parameter $P$ equal to
 $P$ + $\Delta P$ and $P$ - $\Delta P$.
 We note that although the uncertainties in the stellar age are large,
 the derived $p$-value is highly significant (the probability of being
 due to the uncertainties is only 3\%). 
 We also note a slightly larger fraction of 
 F-type stars in the hot-PH sample. As we do not have any explanation for that we will explore this potential bias 
 in more detail in Sect.~\ref{comp_biased}.

\begin{figure}[htb]
\centering
\includegraphics[angle=270,scale=0.45]{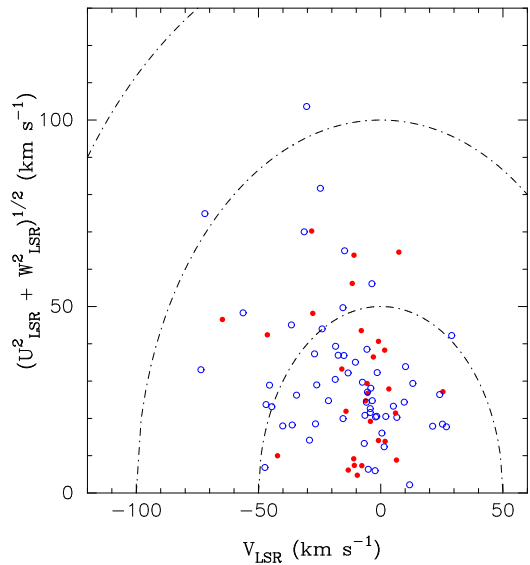}
\caption{
Toomre diagram of the observed stars. Hot PHs are shown by filled red circles while
cool PHs are plotted with open blue circles.
Dash-dot lines indicate constant total velocities,
$ V_{\rm Total} = \sqrt{U^{\rm 2}_{\rm LSR}+ V^{\rm 2}_{\rm LSR} + W^{\rm 2}_{\rm LSR}} =$
50, 100, and 150 km s$^{\rm -1}$.  Median uncertainties for  $V_{\rm LSR}$ and
$V_{\rm Total}$ are of the order of 0.54 km s$^{\rm -1}$. }
\label{toomre_diagram}
\end{figure}

 Regarding kinematics, the kinematic criteria for thick/thin disc membership does not
 reveal any significant difference between hot and cool PHs as most of the stars 
 in both samples appear to belong to the thin disc population. 
 Figure~\ref{toomre_diagram} shows the corresponding Toomre diagram where it can be seen that both hot- 
 and cool-PHs are well mixed in the vertical axis. However, the range of $V_{\rm LSR}$ values of the hot PHs is narrower
 than that of the cool PHs, and therefore a tendency of hot PHs to show larger $V_{\rm LSR}$ values in the
 corresponding cumulative distribution function at low values (between -10 and -50 km s$^{\rm -1}$)
 is found (Fig.~\ref{cummu_bias}, bottom left).
 In addition, a comparison in terms of mean Galactocentric distance, R$_{\rm mean}$,
 was performed as the Galactic birth place has been suggested to be 
 related to the chemical properties of planet hosts \citep{2014A&A...564L..15A,2016A&A...592A..87A}.
 Values of R$_{\rm mean}$ are taken from \cite{2011A&A...530A.138C}.
 While the K-S test reveals no significant differences between % the distributions
 hot and cool PHs, the plot of the corresponding cumulative distributions
 (Fig.~\ref{cummu_bias}, bottom right) shows a tendency of hot PHs to 
 be located at slightly larger R$_{\rm mean}$ values in the R$_{\rm mean}$
 range 6.5-7.8 Kpc.
% (to be checked, Rmean and V related?)

 Other properties such as stellar mass or effective temperature were analysed but
 no differences between the two samples were found. 

% ----------------------------------------------------------------
% Table 4: Biases
% ----------------------------------------------------------------
\begin{table*}[!htb]
\centering
\caption{
Comparison between the properties of cool  and hot PHs.}
\label{bias_table}
%\begin{scriptsize}
\begin{tabular}{llllllllll}
\hline\noalign{\smallskip}
 &  \multicolumn{3}{c}{\textbf{cool PHs}}  & \multicolumn{3}{c}{\textbf{hot PHs}} & \multicolumn{3}{c}{\textbf{K-S test}} \\
 &  \multicolumn{3}{c}{\hrulefill}         & \multicolumn{3}{c}{\hrulefill}       & \multicolumn{3}{c}{\hrulefill}  \\
 &     Mean        &  Median  &   Range       & Mean     & Median  &  Range & $D$    & $p$-value & $p$-sig.  \\
\hline\noalign{\smallskip}
 V (mag)        & 7.3   &       7.5     &       3.7/9.6 &       7.6     & 8.0     &       4.1/10  &       0.32    &    0.03 & 0.31        \\
 Distance (pc)  & 41.6  &       38.11   &       3.2/103 &       43.9    & 42.7    &       12.5/95 &       0.16    &    0.64 & 0.13        \\
 Age (Gyr)      & 4.2   &       3.9     & 0.31/11.4     &  3.0          & 2.4     &  0.1/11       &       0.28    &    0.09 & 0.03        \\
 Mass (M$_{\odot}$) & 1.07 & 1.08       &  0.77/1.41    &  1.08         & 1.07  &  0.83/1.41    &       0.09    &    0.99 & 0.20  \\
 T$_{\rm eff}$ & 5723   & 5761        & 4849/6472       &  5741         & 5656  &  5025/6704    &    0.19       &    0.41 & 0.13 \\
 $V_{\rm LSR}$ (km s$^{\rm -1}$)  & -10.23 & -7.73    &  -64.77/25.45     & -13.49 & -7.49 & -73.53/29.03 & 0.25 & 0.15   & 0.12 \\
 R mean (Kpc)   & 7.43  &  7.53         & 5.97/8.74     & 7.58          & 7.61  &  6.6/8.57     &       0.24    &    0.37 & 0.05     \\
\hline
SpType(\%)         & \multicolumn{3}{l}{7 (F); 74 (G); 19 (K) } & \multicolumn{3}{l}{24 (F); 45 (G); 31 (K)} & \multicolumn{3}{c}{  } \\
D/TD(\%)$^{\dag}$ & \multicolumn{3}{l}{85 (D); 15 (R)} & \multicolumn{3}{l}{93 (D); 7 (R)} & \multicolumn{3}{c}{  } \\
\noalign{\smallskip}\hline\noalign{\smallskip}
\multicolumn{10}{l}{$^{\dag}$ D: Thin disc, TD: Thick disc, R: Transition} \\
%\end{scriptsize}
\end{tabular}
%\end{scriptsize}
\end{table*}

%%------------------------------------------------------
%  Figure 1: distribuciones acumuladas biases
%%------------------------------------------------------

%%------------------------------------------------------
%  Figure 2: Toomre diagram
%%------------------------------------------------------

%+++++++++++++++++++++++++++++++++++++++++++++++++++++++++++++++++++
\section{Analysis}\label{seccion_analysis}
%+++++++++++++++++++++++++++++++++++++++++++++++++++++++++++++++++++
% ----------------------------------------------------------------------
\subsection{Metallicity distributions}\label{metallicity_distributions}
% ----------------------------------------------------------------------

\begin{figure}
\centering
\includegraphics[angle=270,scale=0.45]{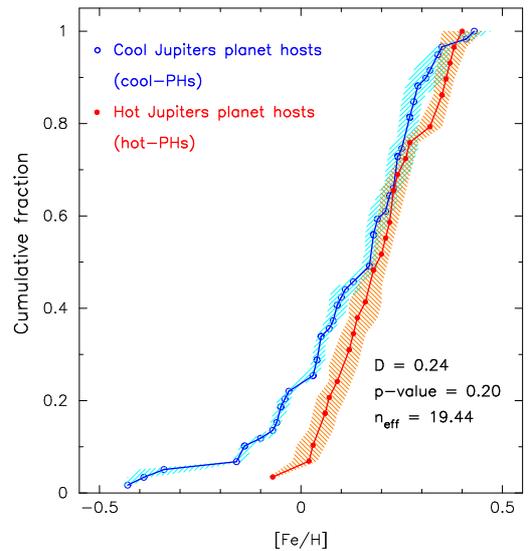}
%hot_cool_metallicity_distribution_ver_feb17.ps§}
\caption{ [Fe/H] cumulative frequencies.} 
\label{distribuciones_acumuladas}
\end{figure}

 The cumulative distribution functions of the metallicity for the
 hot and cool PH samples are presented in Figure~\ref{distribuciones_acumuladas}.
 For guidance, some statistical diagnostics are also given in
 Table~\ref{metal_statistics}.
  The hatched regions indicate the limits of the cumulative distributions
 built by considering that all stars have a metallicity of [Fe/H] + $\Delta$[Fe/H]
 and  [Fe/H] - $\Delta$[Fe/H]. 
 A two-sample K-S test
 shows that the global metallicity distributions of  hot- and cool-PHs are similar
 ($p$-value $\sim$ 20\%, K-S statistics $\sim$ 0.24, n$_{\rm eff}$ $\sim$ 19).
 In order to account for the uncertainties in the derived metallicities, 
 we proceed as in Sect.~\ref{biases}.
 The significance of the original $p$-value is found to be of only
 $\sim$ 6.5\%, thus confirming the reliability of the result. 

 We note that a visual inspection of Figure~\ref{distribuciones_acumuladas} suggests a
 ``deficit'' of hot PHs with respect to cool PHs at metallicity values
between +0.00 and +0.20 dex, even if the uncertainties in the cumulative distributions
 are considered. A similar result was found by
 \citet[][]{2015A&A...579A..20M} but based on a smaller sample
 (only 5 hot PHs and 17 cool PHs).
 In order to find further confirmation of this possible trend,  \citet[][]{2015A&A...579A..20M}, in their Appendix~B, additionally performed a study of the
gas-giant planets listed in  the Extrasolar Planets Encyclopaedia
 \citep{2011A&A...532A..79S}
 and the Exoplanet Orbit Database \citep{2011PASP..123..412W} without applying any
 further selection
 criteria  (and therefore mixing different types of stars and planets
 and with inhomogeneous metallicity determinations).
 The results from this comparison showed a ``deficit'' of hot Jupiters at metallicities between
 -0.50 and +0.00 dex.

 In order to test the significance of this possible deficit we repeated the
 statistical analysis performed before but considering only the stars with
 metallicities below or equal to +0.20 dex. The K-S test returns in this case a K-S statistic
 of 0.37 and a low $p$-value of only $\sim$ 0.08 (slightly larger than the usual
 threshold of 2\% for statistical significant difference). 
 Although the samples' size diminishes significantly, we note that 
 the value of n$_{\rm eff}$ is still large, $\sim$ 10.5  (the K-S test is reasonably accurate for sample sizes for which
 n$_{\rm eff}$ is greater than four\footnote{https://it.mathworks.com/help/stats/kstest2.html} 
 %\citep{1983MNRAS.202..615P} (TO BE CHECHED)
 ), and the significance of the $p$-value returned by our simulations is of the order
 of $\sim$ 3\%. Our conclusion is that the deficit of hot PHs at low metallicities might be real 
 although it may need further confirmation by analysing larger samples. 
 
 We finally note that in our sample there are no hot Jupiters harbouring stars
 with metallicities below $\sim$ -0.10 dex, whilst cool Jupiters
 can be found around more metal-poor stars.

%%%%%%https://it.mathworks.com/help/stats/kstest2.html
 
% We find the metallicity distribution of
% hot-PHs and cool-PHs to be similar. However, the distributions show a 
% ``deficit'' of hot-PHs with respect to cool-PHs at metallicity values
% between +0.00 and +0.20 dex. A similar result was found by
%\citet[][]{2015A&A...579A..20M} but based on a smaller sample
% (only 5 hot-PHs and 17 cool-PHs).
% In order to find further confirmation of this possible trend,  \citet[][]{2015A&A...579A..20M}
% in their Appendix~B, additionally performed an study of the 
% the gas-giant planets listed in  the Extrasolar Planets Encyclopaedia
% \citep{2011A&A...532A..79S}
% and the Exoplanet Orbit Database \citep{2011PASP..123..412W} without applying any
% further selection
% criteria  (and therefore mixing different types of stars and planets
%% different kinds of surveys
% and with inhomogeneous metalliticy determinations).
% The results from this comparison showed a ``deficit'' of hot Jupiters at metallicities between
% -0.50 and +0.00 dex. 

% We note that in our sample there are no hot Jupiters harbouring stars
% with metallicities below $\sim$ -0.10 dex, whilst cool Jupiters
% can be found around more metal-poor stars. 
% A two sample K-S test 
% shows that the metallicity distributions of  hot-PHs and cool-PHs are similar
% ($p$-value 20\%, K-S statistics $\sim$ 0.24, n$_{\rm eff}$ $\sim$ 19).
  
%------------------------------------------------------------------------------
% --- Figura 3: Metallicity distributions
%------------------------------------------------------------------------------

%-------------------------------------------------------------------------------
% --- Table 5: [Fe/H] statistics
%-------------------------------------------------------------------------------
\begin{table}
\centering
\caption{[Fe/H] statistics of the stellar samples.}
\label{metal_statistics}
\begin{scriptsize}
\begin{tabular}{lcccccc}
\hline\noalign{\smallskip}
$Sample$  &  $Mean$ & $ Median$ & $Deviation$  &  $Min$&  $Max$ & $N$ \\
\hline\noalign{\smallskip}
 cool-PHs & 0.12 & 0.18 & 0.18 & -0.43 & 0.43 & 59 \\
 hot-PHs  & 0.20 & 0.20 & 0.12 & -0.07 & 0.40 & 29 \\
\noalign{\smallskip}\hline\noalign{\smallskip}
\end{tabular}
\end{scriptsize}
\end{table}

% ----------------------------------------------------------------------
\subsection{Other chemical signatures}\label{other_chemical_sign}
% ----------------------------------------------------------------------

\begin{figure}[!htb]
\centering
\includegraphics[angle=270,scale=0.45]{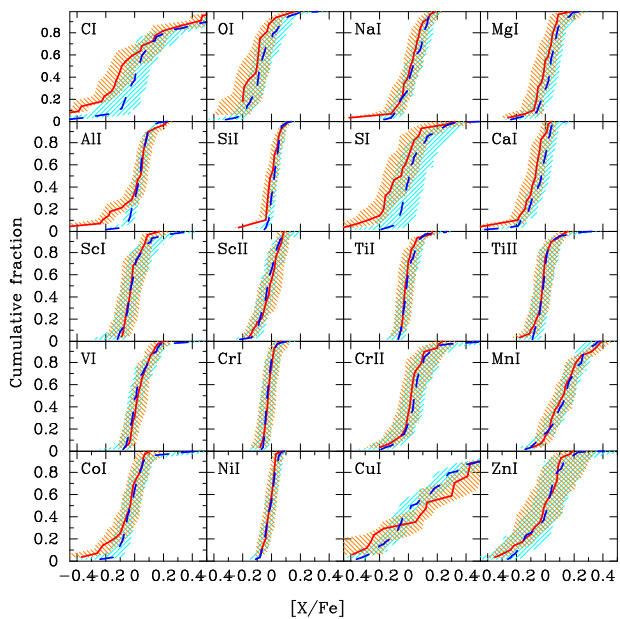}
\caption{ 
[X/Fe] cumulative fraction of hot PHs (red continuous line)
and cool PHs (blue dash-to-dot line).
}
\label{distribuciones_acumuladas_abundancias}
\end{figure}

 In order to find differences in the abundances of other chemical elements besides iron,
 the cumulative distribution [X/Fe]
 \footnote{  Many previous studies dealing with chemical peculiarities in planet hosts use [X/Fe] 
  instead of [X/H] \citep[e.g.][]{2009ApJ...704L..66M,2009A&A...508L..17R,2013A&A...552A...6G,2014A&A...564L..15A,2015A&A...579A..20M}.}
 comparing the abundance distributions of 
 hot- and cool-PHs is shown in Figure~\ref{distribuciones_acumuladas_abundancias}.
 Some statistical diagnostics are also presented in Table~\ref{abundance_statistics},
 where the results of a K-S test for each ion are also listed.
 The tests suggest that there might be differences in the abundances
 of C~{\sc i}, Si~{\sc i}, S~{\sc i}, and Ca~{\sc i} (slightly larger
 abundances for the cool PHs).
  We note that in all these cases our simulations show that the low $p$-values derived
 are not due to the uncertainties in the derived abundances.
 It is worth mentioning that Si~{\sc i} and Ca~{\sc i} are $\alpha$ elements and that
 the $p$-value returned by the K-S test is also very low for Mg~{\sc i},
 another $\alpha$ element.   
 On the other hand, very similar abundance distributions are found
 for Sc~{\sc i}, Cr~{\sc i}, and Mn~{\sc i}, the latter two ions
 being iron-peak elements. %%%\LEt{Please check that I have retained your intended meaning.} 

%%------------------------------------------------------
%  Figure 4: distribuciones acumuladas abundancias
%%------------------------------------------------------

%-------------------------------------------------------------------------------
% --- Table 6: Comparison of abundances
%-------------------------------------------------------------------------------
\begin{table}
\centering
\caption{Comparison between the elemental abundances of
 hot and cool PHs.}
\label{abundance_statistics}
\begin{scriptsize}
\begin{tabular}{lcccccccc}
\hline\noalign{\smallskip}
 [X/Fe] & \multicolumn{2}{c}{Hot PHs}     &  \multicolumn{2}{c}{Cool PHs}   & \multicolumn{4}{c}{K-S test} \\
        &  \multicolumn{2}{c}{\hrulefill} &  \multicolumn{2}{c}{\hrulefill} &  \multicolumn{4}{c}{\hrulefill} \\
        &   Median  &  $\sigma$ & Median  &  $\sigma$ & $D$ & $p$-value     & n$_{\rm eff}$ & $p$-sig. \\
           \hline\noalign{\smallskip}
C/O           &    0.71    &     0.68   &           0.68   &      1.07    &     0.15    &     0.95  &     11.69 & 0.071 \\
Mg/Si         &    1.05    &     0.19   &           1.10   &      0.17    &     0.20    &     0.40  &      19.44 & 0.050 \\
\hline
$X_{\alpha}$       &   -0.02    &     0.05   &           0.00   &      0.05    &     0.42    & $<$ 0.01  &      19.44 & $<$ 0.001 \\
$X_{\rm Fe}$       &    0.02    &     0.06   &           0.02   &      0.06    &     0.14    &     0.81  &      19.44 & 0.128 \\
$X_{\rm vol}$      &   -0.03    &     0.13   &           0.02   &      0.10    &     0.25    &     0.16  &  19.44 & 0.043  \\     
\hline
C~{\sc i}          &   -0.06    &     0.31   &           0.04   &      0.28    &     0.40    &     0.01  &      16.02 &  0.002 \\
O~{\sc i}          &   -0.09    &     0.10   &          -0.08   &      0.12    &     0.30    &     0.17  &      12.54 &  0.018 \\
Na~{\sc i}         &    0.02    &     0.11   &           0.04   &      0.08    &     0.20    &     0.37  &      19.44 &  0.047 \\
Mg~{\sc i}         &    0.00    &     0.08   &           0.04   &      0.08    &     0.28    &     0.08  &      19.44 &  0.008 \\
Al~{\sc i}         &    0.05    &     0.15   &           0.04   &      0.07    &     0.20    &     0.41  &      18.88 &  0.076 \\
Si~{\sc i}         &    0.00    &     0.06   &           0.02   &      0.04    &     0.33    &     0.02  &      19.44 &  0.003 \\
S~{\sc i}          &   -0.04    &     0.17   &           0.01   &      0.14    &     0.43    & $<$ 0.01  &      18.55 & $<$ 0.001 \\
Ca~{\sc i}         &   -0.06    &     0.10   &          -0.02   &      0.09    &     0.35    &     0.01  &      19.44 &  0.002 \\
Sc~{\sc i}         &   -0.02    &     0.07   &          -0.02   &      0.10    &     0.11    &     0.98  &      17.28 &  0.056 \\
Sc~{\sc ii}        &   -0.01    &     0.07   &          -0.02   &      0.07    &     0.14    &     0.79  &      19.44 &  0.081 \\
Ti~{\sc i}         &   -0.02    &     0.05   &          -0.01   &      0.06    &     0.16    &     0.64  &      19.44 &  0.085 \\
Ti~{\sc ii}        &   -0.01    &     0.07   &          -0.01   &      0.08    &     0.15    &     0.75  &      19.44 &  0.086 \\
V~{\sc i}          &    0.03    &     0.07   &           0.02   &      0.10    &     0.16    &     0.63  &      19.44 &  0.074 \\
Cr~{\sc i}         &   -0.02    &     0.03   &          -0.02   &      0.04    &     0.11    &     0.96  &      19.44 &  0.044 \\
Cr~{\sc ii}        &    0.02    &     0.09   &           0.05   &      0.12    &     0.28    &     0.07  &      19.44 &  0.004 \\
Mn~{\sc i}         &    0.13    &     0.13   &           0.14   &      0.12    &     0.13    &     0.90  &      19.44 &  0.112 \\
Co~{\sc i}         &   -0.02    &     0.11   &          -0.03   &      0.11    &     0.16    &     0.65  &      18.99 &  0.088 \\
Ni~{\sc i}         &   -0.01    &     0.04   &           0.00   &      0.04    &     0.14    &     0.84  &      19.44 &  0.091 \\
Cu~{\sc i}         &    0.13    &     0.34   &           0.06   &      0.28    &     0.26    &     0.27  &      13.15 &  0.034 \\
Zn~{\sc i}         &    0.00    &     0.14   &           0.02   &      0.15    &     0.13    &     0.89  &      19.44 &  0.080 \\         
\noalign{\smallskip}\hline\noalign{\smallskip}
\end{tabular}
\end{scriptsize}
\end{table}

% We thereferore grouped the ions into three categories:
 We therefore grouped the ions into three categories: alpha elements,
 iron-peak elements,
 and volatile elements. Following the definitions provided in \cite{2014A&A...566A..83M} and \cite{2017A&A...602A..38M}
 we considered  Mg, Si, Ca, and Ti as alpha elements; Cr, Mn, Co, and Ni as iron-peak elements; and
 C, O, Na, S, and Zn  as volatiles. 

\begin{figure*}[!htb]
\centering
\begin{minipage}{0.32\linewidth}
\includegraphics[angle=270,scale=0.38]{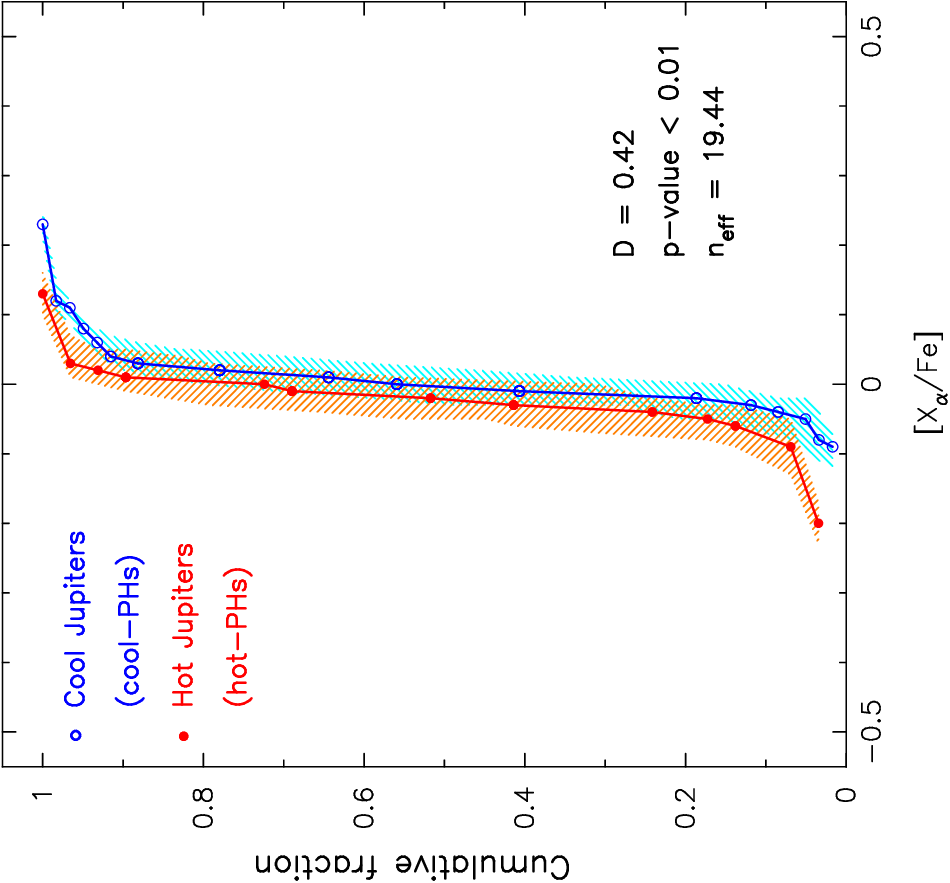}
\end{minipage}
\begin{minipage}{0.32\linewidth}
\includegraphics[angle=270,scale=0.38]{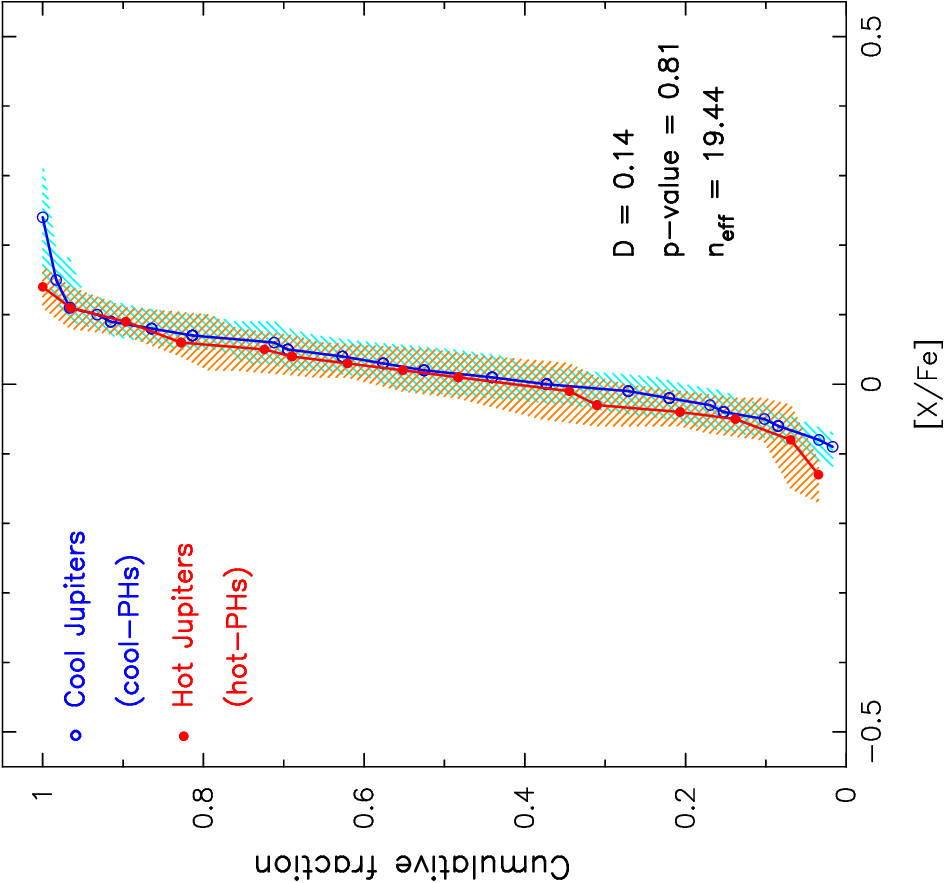}
\end{minipage}
\begin{minipage}{0.32\linewidth}
\includegraphics[angle=270,scale=0.38]{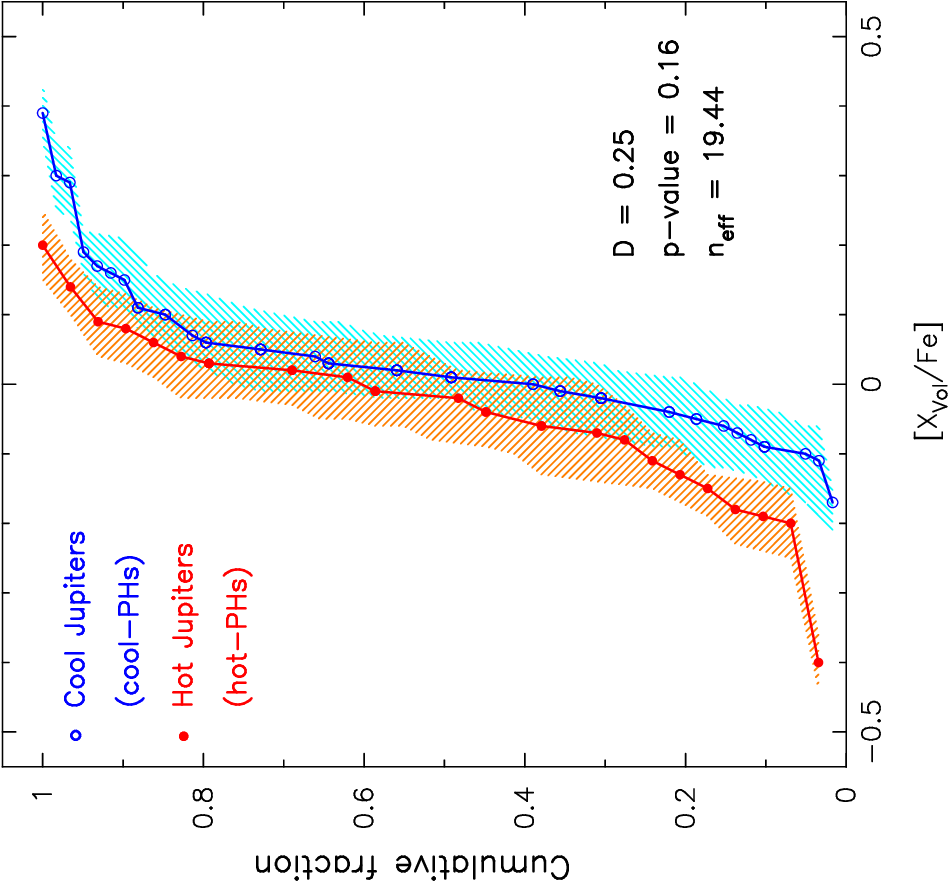}
\end{minipage}
\caption{ 
 Histogram of cumulative frequencies of [X$_{\alpha}$/Fe] (left),
 [X$_{\rm Fe}$/Fe] (middle), and [X$_{\rm vol}$/Fe] (right).}
\label{distrib_acu_alpha_iron_vol}
\end{figure*}

 The corresponding cumulative functions are shown in Figure~\ref{distrib_acu_alpha_iron_vol},
 where [X$_{\alpha}$/Fe], [X$_{\rm Fe}$/Fe], and [X$_{\rm vol}$/Fe] denote
 the abundances of alpha, iron-peak, and volatile elements, respectively.
 The results from the K-S test show no abundance difference in iron elements
 between hot- and cool PHs. However, the test suggests significant differences
 when $\alpha$ elements are considered, showing cool PHs to have slightly larger abundances.
  For volatile elements, the K-S test returns a probability of $\sim$ 16\% that both samples
 have similar distributions, not low enough to claim a
 statistical significant difference. 
  Possible age effects affecting these results are discussed in Sect.~\ref{comp_biased}.

%%---------------------------------------------------------------------------
%  Figure 5. Cumulative frequencies of alpha, iron-peak and volatile
%%---------------------------------------------------------------------------

\begin{figure*}[!htb]
\centering
\begin{minipage}{0.49\linewidth}
\includegraphics[angle=270,scale=0.45]{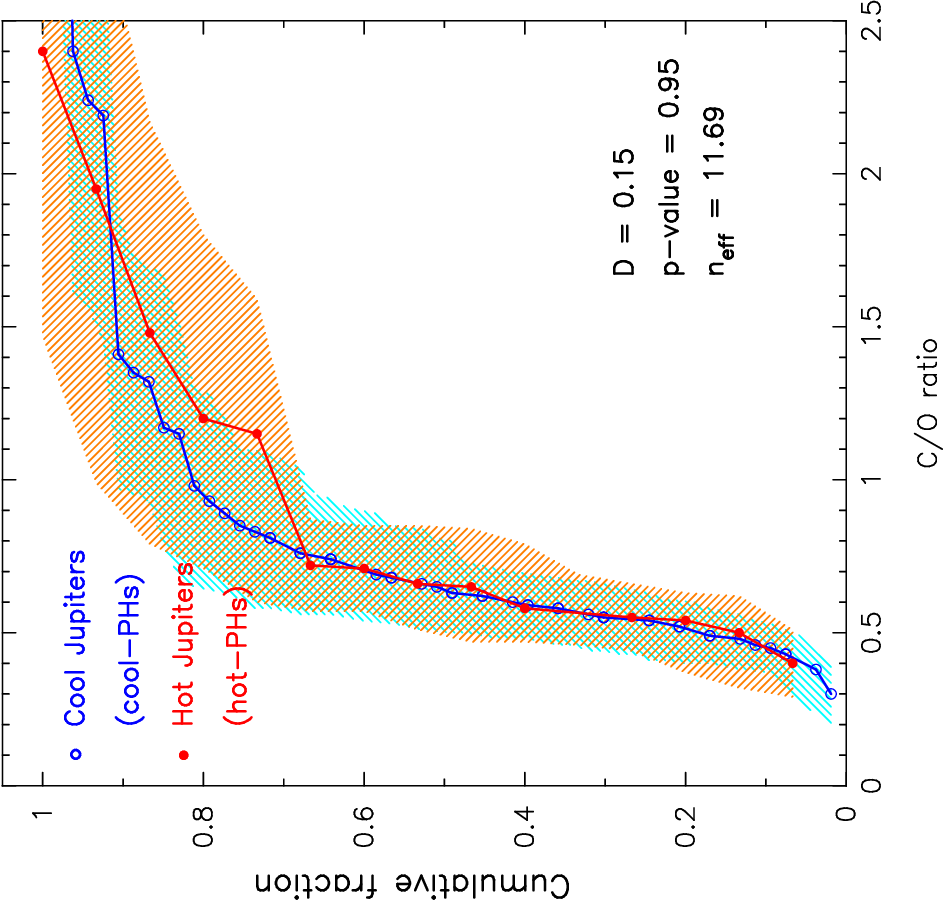}
%hc_stars_distribucion_COratio.ps}
\end{minipage}
\begin{minipage}{0.49\linewidth}
\includegraphics[angle=270,scale=0.45]{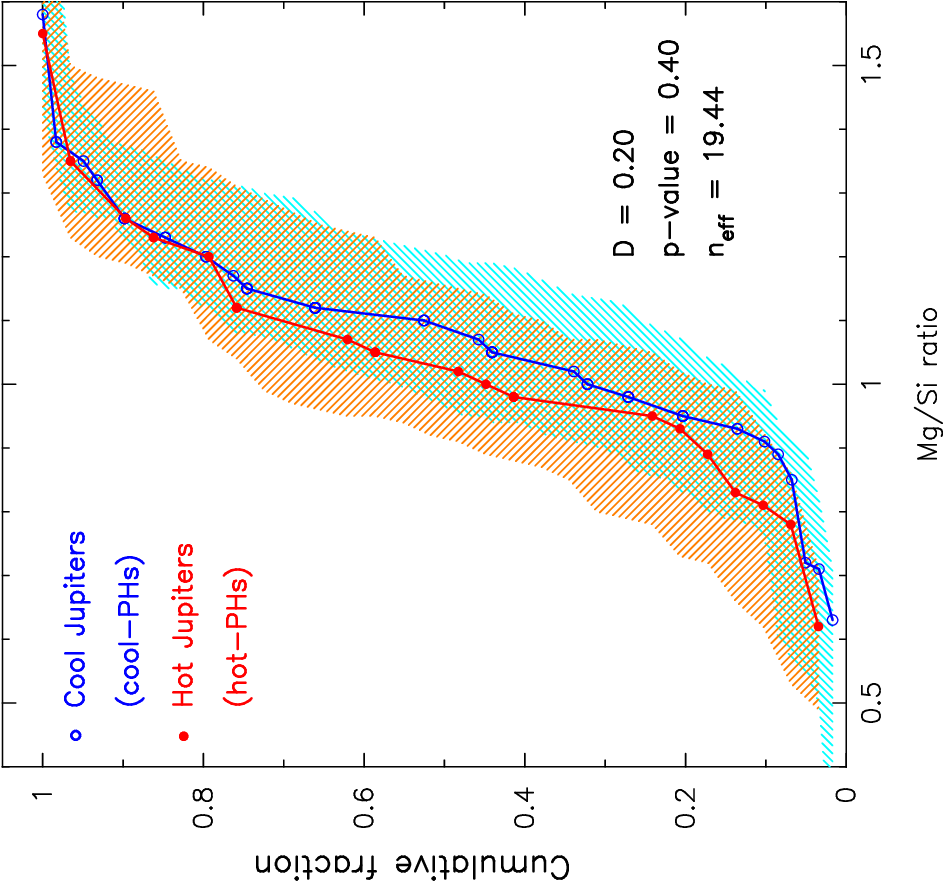}
%hc_stars_distribucion_MgSiratio.ps}
\end{minipage}
\caption{
 Histogram of cumulative frequencies of C/O (left), and Mg/Si (right).}
\label{distrib_acu_co_mg_ratios}
\end{figure*}

 Finally, the stellar C/O and Mg/Si ratios were also considered; see Figure~\ref{distrib_acu_co_mg_ratios}. 
 These ratios are known to play an important role in constraining the
 planetary composition \citep[e.g.][]{2010ApJ...715.1050B,2015A&A...580A..30T,2015A&A...577A..83D}.  
  No statistically significant differences between hot and cool PHs
 have been found.

%The corresponding plots show that hot-PHs tend to have slightly larger C/O ratios
% while larger Mg/Si ratios might be found in cool-PHs. However, neither of these
% tendencies seem to be statistically significant. 

%%---------------------------------------------------
%  Figure 6: C/O and Mg/Si ratios
%%---------------------------------------------------

% ----------------------------------------------------------------------
\subsection{Stellar metallicity and planetary properties}\label{planetary_stellar}
% ----------------------------------------------------------------------

%%------------------------------------------------------
%  Figure 7:  [Fe/H] vs periodos y eccentricidades
%%------------------------------------------------------
\begin{figure*} %[!htb]
\centering
\begin{minipage}{0.49\linewidth}
\includegraphics[angle=270,scale=0.55]{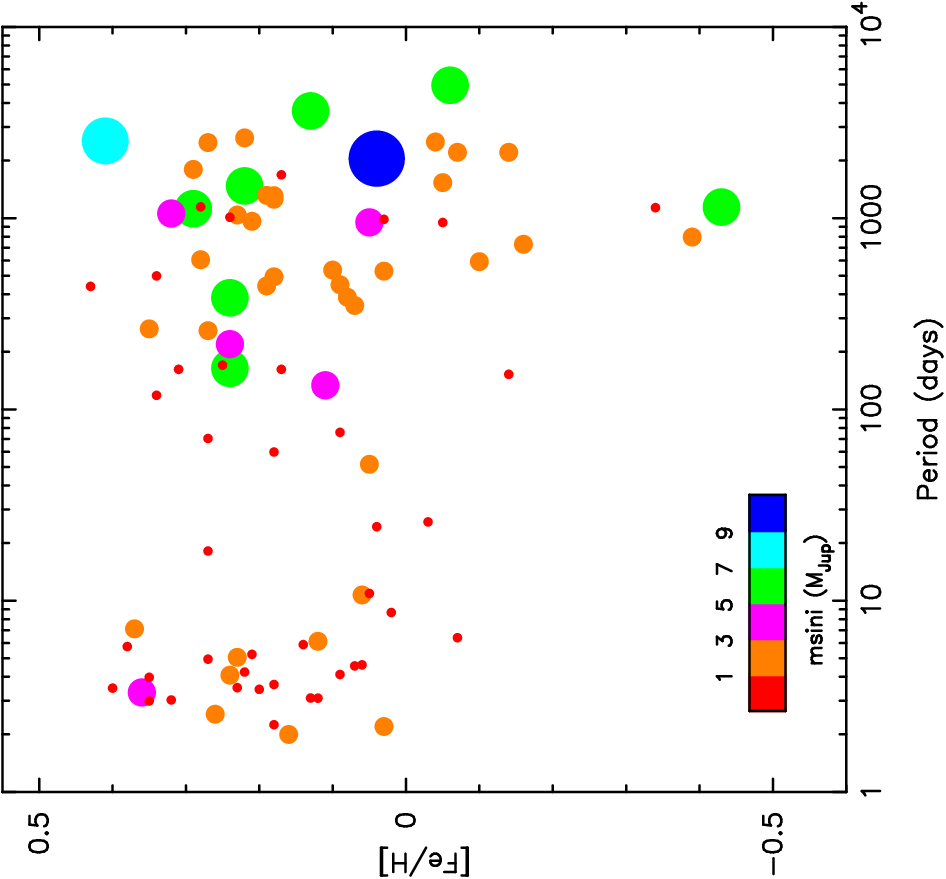}
%%period-masa-diagram-metal2_rev.ps}
%period-masa-diagram-metal2.ps}
\end{minipage}
\begin{minipage}{0.49\linewidth}
\includegraphics[angle=270,scale=0.55]{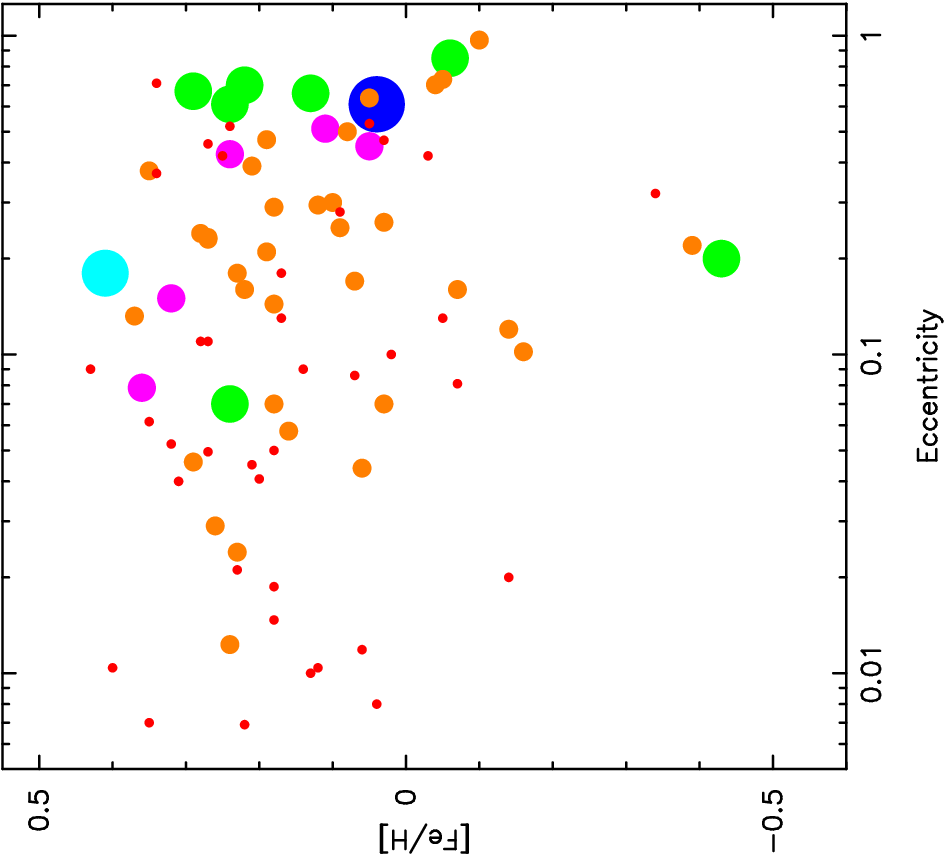}
\end{minipage}
\caption{
 [Fe/H] as a function of the orbital period (left panel) and eccentricity (right panel).
 Stars are plotted with different colours and increasing sizes with increasing  planet minimum mass.
  Median uncertainties are 0.6 M$_{\rm Jup}$ for the planetary minimum mass, 0.6 days for the planetary
 periods, 0.04 for the eccentricities, and 0.02 dex for [Fe/H].}
\label{pmmetal_diagram}
\end{figure*}

  We perform a search for correlations between the stellar abundances
  and the planet orbital properties. Figure~\ref{pmmetal_diagram} shows the stellar
  metallicity as a function of the orbital period (left) and versus the
  planetary eccentricity (right). Colours and symbols, indicate different
  planetary masses. % (right). %% and periods, respectively.
  The Figure shows  a lack of massive planets at short orbital periods.
  We note that all planets (except one) with m$_{\rm p}\sin i$ larger that 3 M$_{\rm Jup}$
  have a period larger than 100 days. Indeed, most of the planets with
  m$_{\rm p}\sin i$ larger than 1 M$_{\rm Jup}$ have periods larger than 100 days.
  The Figure also shows that planets at short periods have larger metallicities
  (in agreement with the results from Sect.~\ref{metallicity_distributions}).
  All stars (except 2) with planets with $P$ $<$ 100 days show metallicities larger than + 0.00 dex.
 % a tendency of more distant planets (larger periods) to
 % have higher masses and a wider range of metallicities
 % (in agreement with the results from Sect.~\ref{metallicity_distributions}).
 % In particular we note that all planets (except one) with m$_{\rm p}\sin i$ larger that 3 M$_{\rm Jup}$
 % have a period larger than 100 days. Indeed, most of the planets with
 % m$_{\rm p}\sin i$ larger than 1 M$_{\rm Jup}$ have periods larger than 100 days.
 % All stars with planets with $P$ $<$ 100 days but two
 % show metallicities larger than + 0.00 dex.  
  A tendency of higher eccentricities towards more massive planets also seems to
  be present in the data.
  Most of the planets with  m$_{\rm p}\sin i$ larger that 3 M$_{\rm Jup}$ %periods larger than 100 days
  show eccentricities larger than 0.1. 

% ---------------------------------------------------
\subsection{Comparison with previous works}
% ---------------------------------------------------

  The absence of massive planets orbiting at short periods around low-metallicity stars
  in radial velocity surveys was already noted by \cite{2002A&A...390..267U,2003A&A...398..363S,2005ApJ...622.1102F}.
   In a recent work, \cite{2017MNRAS.466..443J} discuss a tendency of stars
  with short-period gas giant planets to have higher metallicities than stars
  with planets at longer periods.
  Our results also agree with
  published results based on the analysis of the transit survey planets 
   \citep[e.g.][]{2013ApJ...763...12B} where  the smallest {\sc Kepler}
  planetary candidates orbiting metal-poor stars were found to show
  a possible period dependency: small planets orbiting metal-poor stars are located at large periods while
  small planets closer to the star tend to have higher metallicities.
   On the other hand, the presence of massive gas-giant planets at long orbital periods 
  (even around low-metallicity stars)
  might be explained if they are mainly formed by gravitational instabilities in the disc
  \citep[e.g.][]{2009ApJ...707...79D}, a mechanism
  that does not depend on the metallicity of the primordial disc %(Boss 1997, 2002, 2006)
 \citep{1997Sci...276.1836B,2002ApJ...567L.149B,2006ApJ...643..501B}.

 \cite{2013ApJ...767L..24D} find that gas-giant planets at short orbital
 distances ($a$ $<$ 0.1 au) around metal-poor stars
 are confined to lower eccentricities while eccentric proto-hot Jupiters undergoing tidal
 circularization orbit metal-rich stars. 
 A comparison with this work   %\cite{2013ApJ...767L..24D}
  seems  complicated given the relatively low number of planet hosts
  in our sample that are located in the
  so-called Period Valley, 0.10 - 1.00 au semimajor axis range
  or 10 days $<$ $P$ $<$ 100 days \citep[e.g.][]{2003MNRAS.341..948J}.
  For this purpose the period-eccentricity diagram of our
  sample is shown in the upper left panel of Figure~\ref{period_eccc_abun} where stars are plotted
  with different colours according to their metallicities.
   It is clear from this Figure that our planet hosts are concentrated
    in two different regions of the plot: one of planets with
     periods $P$ $<$ 10 days and eccentricities lower than 0.1 (with 25 stars);
      and another of periods $P$ $>$ 50 days and eccentricities larger
       than $\sim$ 0.05 (with 52 planet-hosts).
        The ``period valley'' is also visible in our sample (although showing a narrower gap) with only
   three planets with periods between 10 and 50 days.

  Our results seem to agree qualitatively with \cite{2013ApJ...767L..24D}.
  First, for small planets located close to the parent stars
  (m$_{\rm p}\sin i$ $<$ 1 M$_{\rm Jup}$, $P$ $<$ 10 days),
  planet-hosts show positive
  metallicities and their planets have low eccentricities. %%%% large eccentricities $\sim$ 1.
  As in \cite{2013ApJ...767L..24D} hot Jupiters are found around metal-rich stars. 
  For the stars in the valley, \cite{2013ApJ...767L..24D} find mostly metal-rich planet hosts, with a large
  range of eccentricities.  
  In this period range our sample contains three  stars with
  positive metallicities, two hosting planets at high % however, we note that the planets orbiting them both show high
  eccentricity ($e$ $>$ 0.6) and the other one at low eccentricity ($e$ $<$ 0.01).  
  Finally, as in \cite{2013ApJ...767L..24D}, for large planetary periods, %%%for $a$ $>$ 0.1 au,
  low- and high-metallicity stars are mixed covering a large range of eccentricities. 

  Our findings are also in agreement with \cite{2013A&A...560A..51A} who showed that very massive
  planets with eccentric orbits have longer periods than those with
  circular orbits and also that most of the planets with long periods
  orbit around low-metallicity stars.
  In more recent works,
  \cite{2017arXiv170107654B} and \cite{2017A&A...603A..30S}
  state that a transition in metallicity at m$_{\rm p}\sin i$ $\sim$ 5-4 M$_{\rm Jup}$ may occur,
   the stars hosting massive gas-giant planets being on average more metal-poor %. on average
  than the stars hosting planets with masses below 5-4 M$_{\rm Jup}$.
%  {\bf A possible correlation between planetary mass and stellar metallicity have been 
%  also suggested by \cite{2017MNRAS.466..443J}. }
  While a tendency of lower metallicities towards higher planetary masses seems to be
  present in our data, as previously suggested by \cite{2007A&A...464..779R}  and \cite{2017MNRAS.466..443J},
  we do not find any significant metallicity difference between
  planet hosts with masses below and above 4 M$_{\rm Jup}$
  (a K-S test returns a probability $\sim$ 99\% of planet hosts with masses below and above
  4 M$_{\rm Jup}$ to have similar metallicity distributions).
  However, it should be noted that our sample and the one discussed in \cite{2017A&A...603A..30S}
  differ in selection criteria, size, and range of considered planetary masses. %%% properties. 
  
  Differences in the eccentricity distribution of planets with masses above and
  below 4 M$_{\rm Jup}$ have also been reported.
  \cite{2007A&A...464..779R} found that massive planets have an eccentricity distribution consistent with that of 
  binary systems. On the other hand, less massive planets have lower eccentricities.
  A mild tendency of higher eccentricities towards higher planetary masses seems
  consistent with our data.
  We performed a K-S on the eccentricity distribution of planets with 
  masses above and below 4 M$_{\rm Jup}$.
  The results confirm the larger eccentricity values of the massive planets
  ($D$=0.35, $p$-value = 0.05, n$_{\rm eff}$ = 14.2) in agreement with \cite{2007A&A...464..779R}.

% -------------------------------------------------------
\subsection{Comparison with brown dwarf results}
% -------------------------------------------------------

   The findings of the previous subsection
  can be compared with the results from stars harbouring
  companions in the brown dwarf regime.
  \cite{2017A&A...602A..38M} found that stars harbouring brown dwarfs with (minimum) masses
  M$_{\rm C}$$\sin i$ $<$ 42.5 M$_{\rm Jup}$
  tend to have slightly larger metallicities and lower eccentricities than
  stars harbouring brown dwarfs in the range M$_{\rm C}$$\sin i$ $>$ 42.5 M$_{\rm Jup}$.
  It is interesting to note that hot-PHs, such as stars harbouring the less massive brown dwarfs,
  tend to have slightly larger metallicities, lower eccentricities, and less massive companions
  (when compared with cool-PHs). 
  This could suggest a common  formation mechanism for hot Jupiters and low-mass 
  brown dwarfs in which metallicity plays a significant role;
  that is, the core-accretion model. 
 % role as it could be the core-accretion model.
  On the other hand, for cool Jupiters and massive brown dwarfs a metallicity
  dependency does not appear in the data and thus the metal content might not be involved in their formation mechanism. 
  Despite the similarities in metallicity and eccentricity, no further analogies  between brown dwarfs and gas-giant planets
  are found. In particular,  the different period distributions should be noted. Only one brown dwarf in the
  \cite{2017A&A...602A..38M} sample has a period shorter than 10 days.

%+++++++++++++++++++++++++++++++++++++++++++++++++++++++++++++++++++++
\subsection{Other chemical signatures and the period-eccentricity distribution} 
%++++++++++++++++++++++++++++++++++++++++++++++++++++++++++++++++++++++

\begin{figure*} %[!htb]
\centering  %period-eccentricity-metallicity.ps
\begin{minipage}{0.32\linewidth}
\includegraphics[angle=270,scale=0.38]{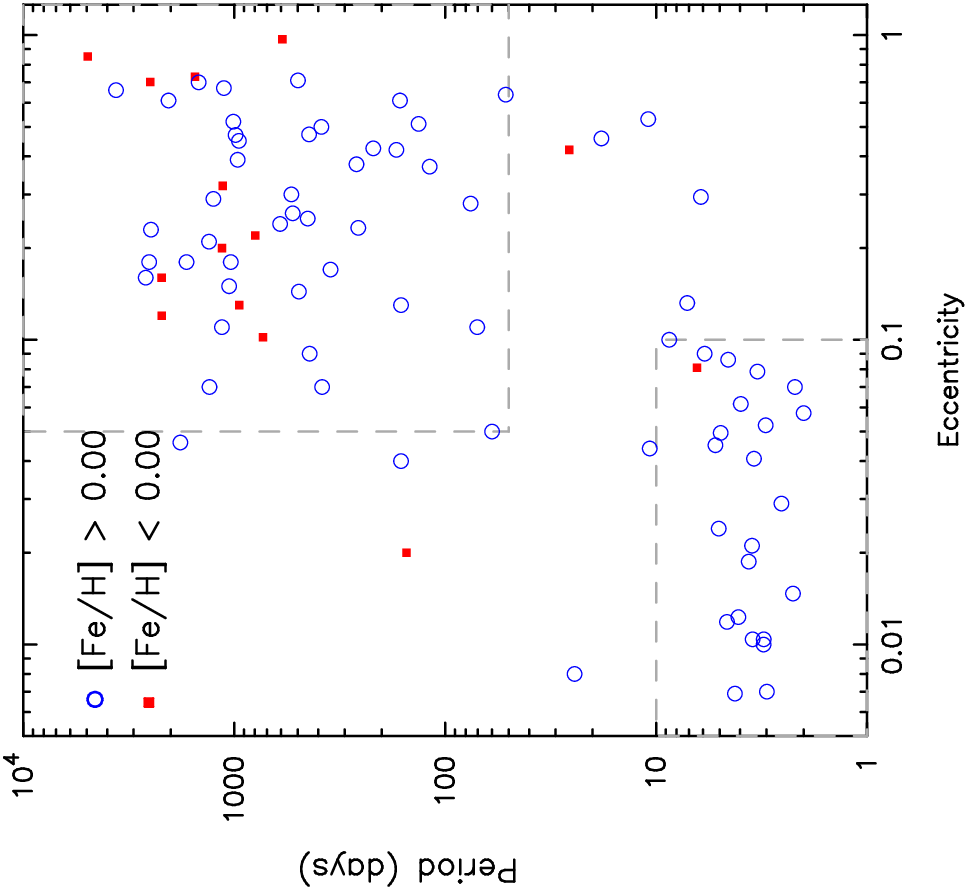}
\end{minipage}
\begin{minipage}{0.32\linewidth}
\includegraphics[angle=270,scale=0.38]{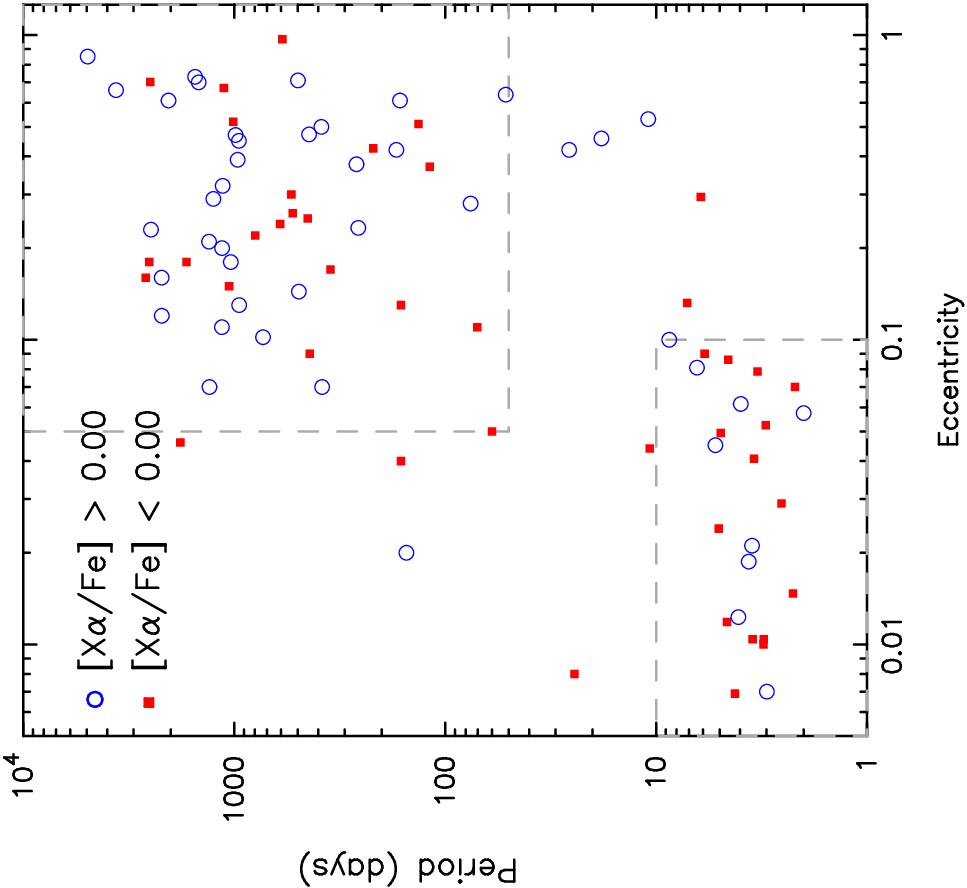}
\end{minipage}
\begin{minipage}{0.32\linewidth}
\includegraphics[angle=270,scale=0.38]{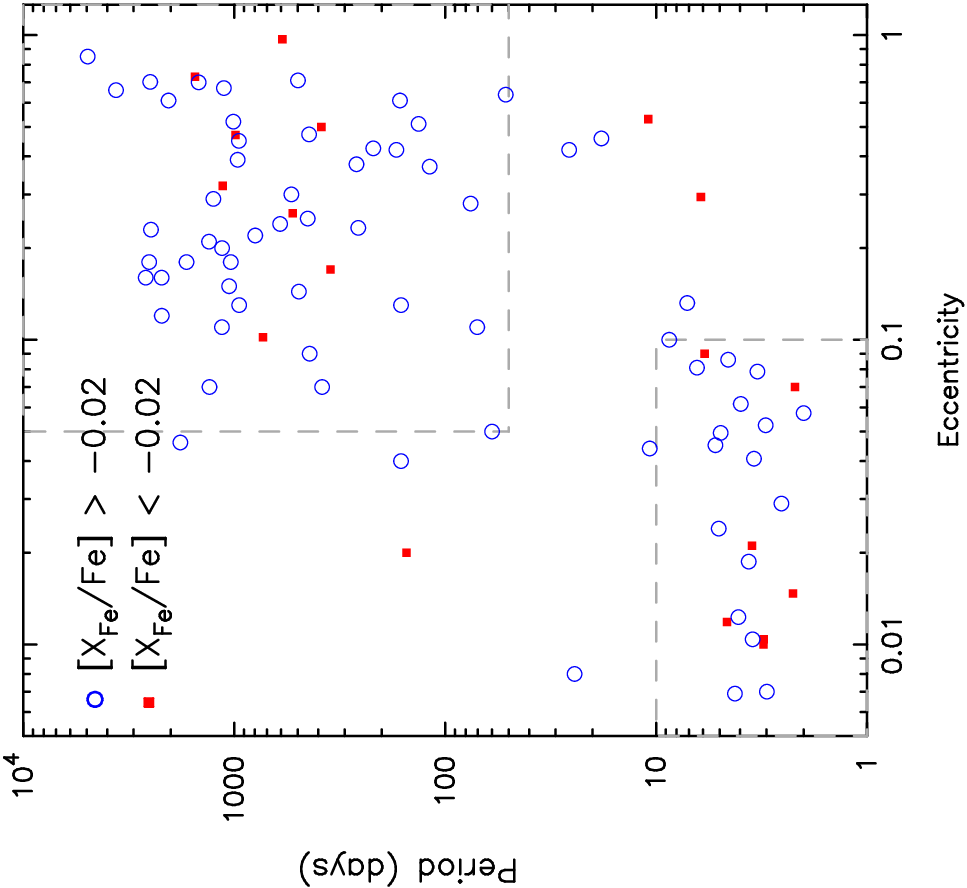}
\end{minipage}
\begin{minipage}{0.32\linewidth}
\includegraphics[angle=270,scale=0.38]{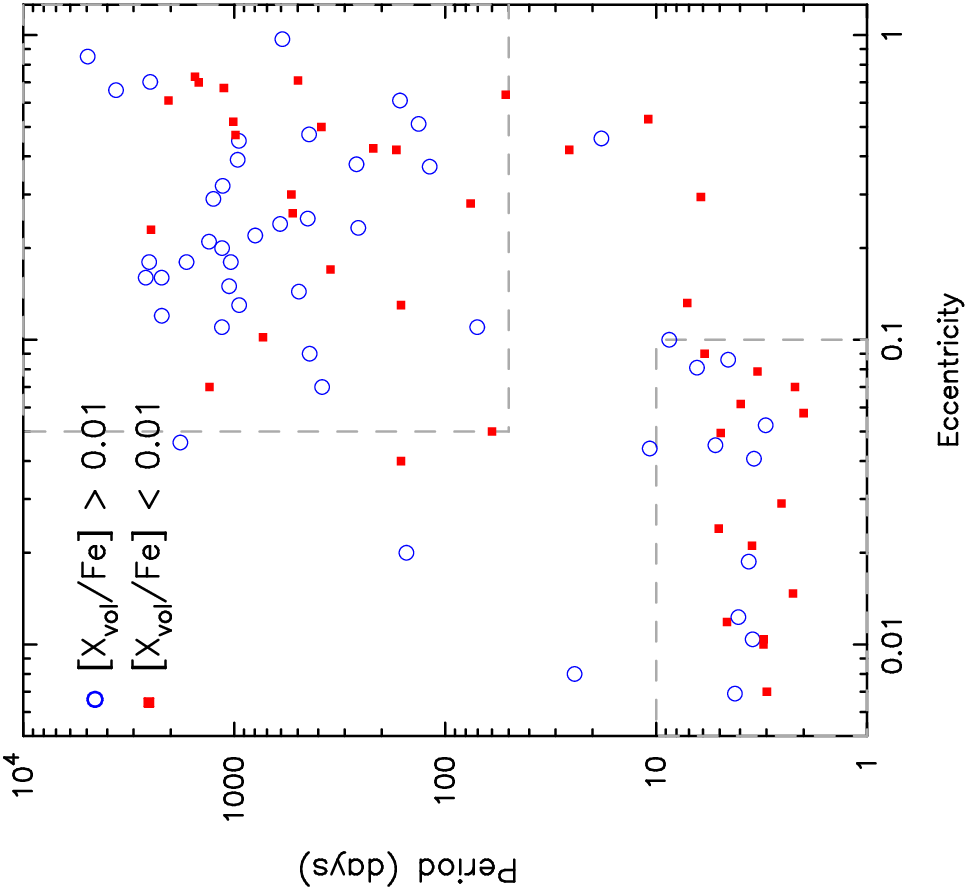}
%period-eccentricity-xvol.ps}
\end{minipage}
\begin{minipage}{0.32\linewidth}
\includegraphics[angle=270,scale=0.38]{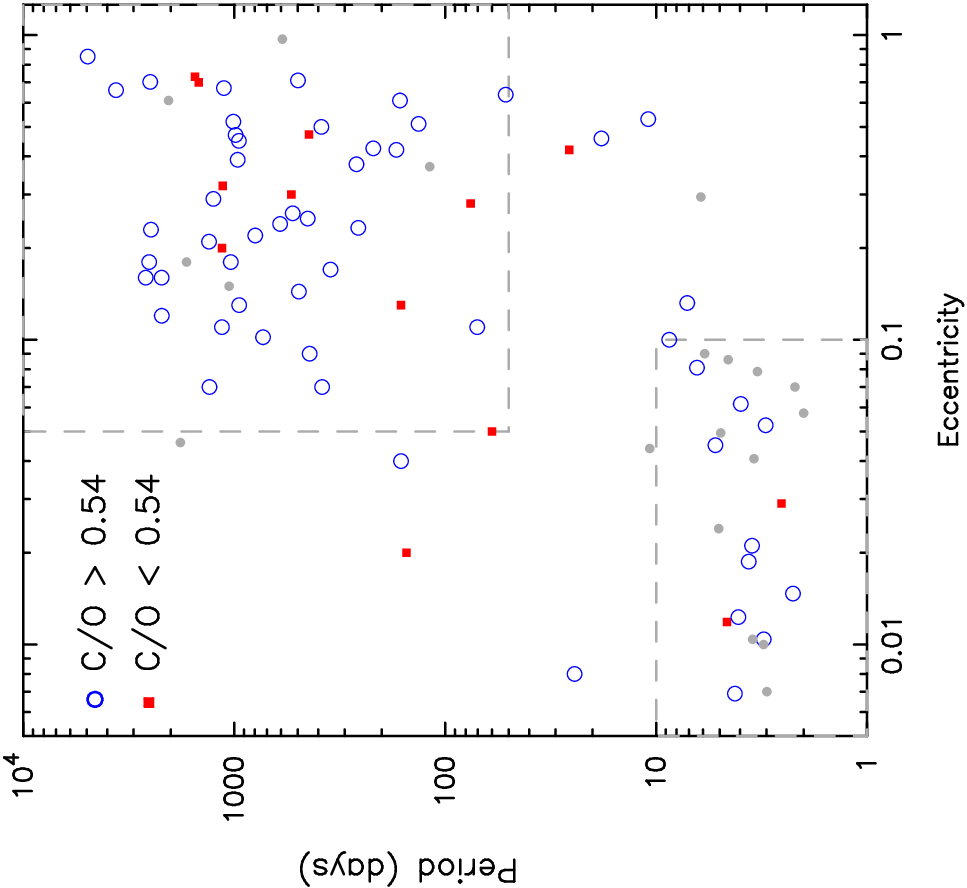}
%period-eccentricity-coratio.ps}
\end{minipage}
\begin{minipage}{0.32\linewidth}
\includegraphics[angle=270,scale=0.38]{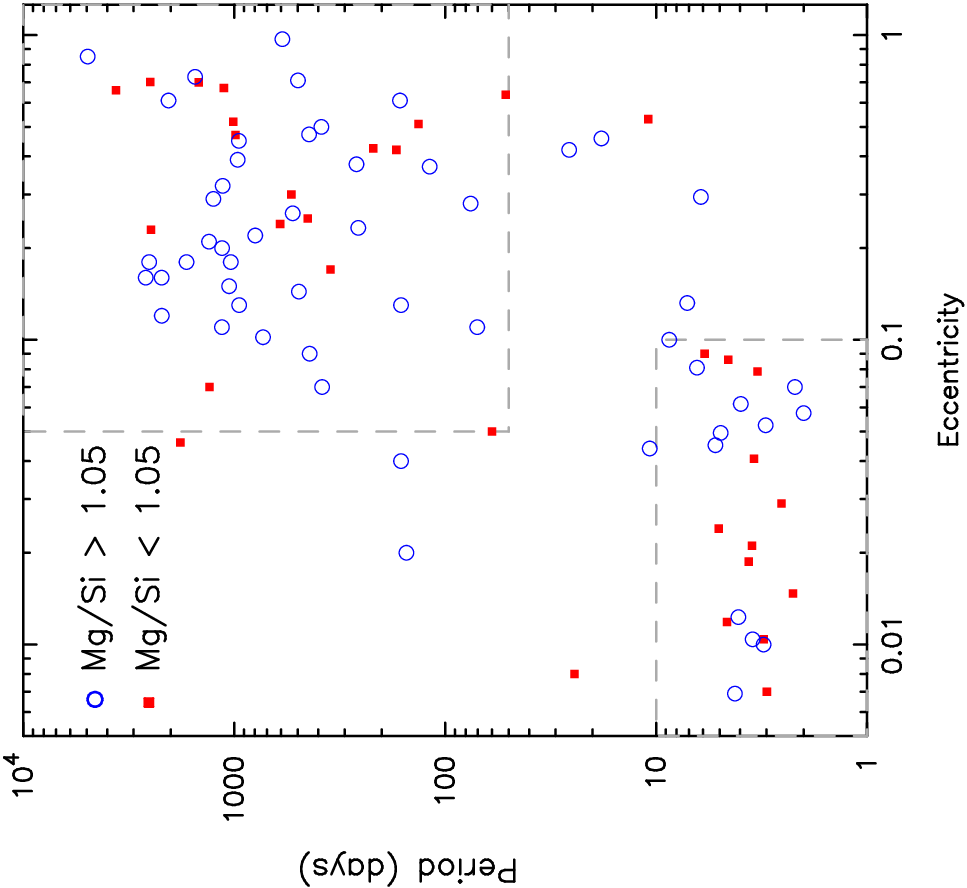}
\end{minipage}
\caption{
 Period vs. eccentricity diagram. Stars are plotted with different colours and
 symbols according to their corresponding  [Fe/H] (top left), X$\alpha$ (top middle), X$_{\rm Fe}$ (top right) 
 X$_{\rm vol}$ (bottom left), C/O (bottom middle), and Mg/Si (bottom right) values.
 Typical uncertainties are 0.6 days (period), 0.04 (eccentricity), 0.02 dex (metallicity, X$\alpha$, and X$_{\rm Fe}$),
 0.05 dex (X$_{\rm vol}$), 0.15 (C/O ratio), and 0.10  (Mg/Si ratio). 
}
\label{period_eccc_abun}
\end{figure*}

 We finally explore the relationships between the planetary position in 
 the period-eccentricity diagrams and the stellar abundances as quantified by the  
 [X$_{\alpha}$/Fe], [X$_{\rm Fe}$/Fe],
 [X$_{\rm vol}$/Fe] values, as well as the stellar C/O and Mg/Si ratios.
 The stars are divided into stars with  low- and high-[X/Fe] values.
  Table~\ref{tabla_period_ecc} provides the percentage of low-abundance stars
 in the population of short and long period planets for each considered abundance.
 In order to do that we first estimate the [X/Fe] distribution of non-planet host stars.
 The spectroscopic abundances provided by \cite{2015A&A...579A..20M}
 were used as they were computed using the same methods and similar
 spectra to those used in this work. %%%Stars with planets were not considered.
 In addition, a 3$\sigma$ clipping procedure was done to remove outliers.
 The corresponding median abundances are
 0.00, -0.02,  0.01, 0.54, and 1.05, respectively, for [X$_{\alpha}$/Fe], [X$_{\rm Fe}$/Fe],
 [X$_{\rm vol}$/Fe], C/O, and
 Mg/Si. 
 Figure~\ref{period_eccc_abun} shows the corresponding period-eccentricity
 plots. % We focus in this plot as we have already seen the period dependence on
% the planetary mass. 
 Stars are plotted with different colours and symbols depending on
 whether their [X/Fe] values are below or above the median values derived
 from the \cite{2015A&A...579A..20M} distributions. 
% {\bf Table~\ref{tabla_period_ecc} provides the percentage of low-abundance stars
% in the population of short and long period planets for each considered abundance.}

 For each considered abundance ([X$_{\alpha}$/Fe], [X$_{\rm Fe}$/Fe],
 [X$_{\rm vol}$/Fe], C/O, and Mg/Si),
 the percentage of stars with
 low and high abundances in the population of short- and long-period planets
 was computed. 
 In order to test the significance of the derived percentages a series of
 10$^{\rm 6}$ simulations was performed.
 In each simulation each star was given a new abundance value chosen randomly
 assuming an underlying Gaussian abundance distribution with the
 same parameters (mean and sigma) as those derived from the
 \cite{2015A&A...579A..20M} data  and the percentage of low-abundance
 stars was computed.
 Assuming that the distribution of the simulated percentages of low-abundance stars
 follows a Gaussian distribution we then compute the probability that
 the simulated percentage takes the value (within 5\%) found when analysing the original
 data. The results are
 given in Table~\ref{tabla_period_ecc}.

 When considering $\alpha$-elements, the low-abundance stars tend to
 dominate the short-period part of the diagram, in agreement with
 the results from Sect.~\ref{other_chemical_sign}.
 The percentage of low $\alpha$ abundance stars in the
 short-period, low-eccentricity part of the diagram is
  $\sim$ 64\%. However, in the long-period, large-eccentricity
 region, stars with low and high $\alpha$ abundances are well mixed,
 and the percentage of low-$\alpha$ stars decreases to 40\%.
 The simulations show that the probability of obtaining  a distribution with 64\% low-$\alpha$ stars in a sample of 25 stars by chance is only $\sim$ 4\%.
 On the other hand, for the long-period, high-eccentricity subsample,
 the probability of getting 40\%  low-$\alpha$ stars in a
 sample of 52 stars is high, of  18.5\%, consistent
 with a random mixture of high and low $\alpha$ abundances.

 A similar analysis was performed for iron-peak and volatile elements,
 as well as for the C/O and Mg/Si ratios. The results are
 given in Table~\ref{tabla_period_ecc}.
 No trends with the iron-peak,  volatiles, or C/O, and Mg/Si ratios seem to be present
 in the data. %% (and therefore, they are not shown in Figure~\ref{period_eccc_abun}). 
  We note that a visual inspection of 
 the C/O in Figure~\ref{period_eccc_abun}
 reveals only two low-abundance stars in the short-period planets group.
 However, the percentage of low C/O stars in this group is $\sim$ 15.4\%,
 similar to the one in the long period planets group ($\sim$ 19.2\%).

%Regarding the volatile elements, we note that the fraction of low-volatile abundance
% stars in the short-period, low-eccentricity planet-hosts subsample
% is  $\sim$ 80\% and highly significant (probability of being
% by chance of only 0.13\%). In contrast, for planet hosts at larger periods
% and eccentricities, low and high-volatile abundance stars are well mixed,
% see Figure~\ref{period_eccc_abun}, bottom right panel. 

%%------------------------------------------------------
%  Figure 8: period-eccentricity diagram
%%------------------------------------------------------

%--------------------------------------------------------------------------------
%  --- Table 7: Fraction of low-abundance stars
%-------------------------------------------------------------------------------
\begin{table}
\centering
\caption{
 Fraction of low-abundance stars in the period-eccentricity diagram.
 Each value is accompanied by its corresponding probability ``by chance''.
}
\label{tabla_period_ecc}
\begin{scriptsize}
\begin{tabular}{l|cc|cc}
\hline\noalign{\smallskip}
         & \multicolumn{2}{c}{$P$ $<$ 10 days, e $<$ 0.1}  &  \multicolumn{2}{c}{$P$ $>$ 50 days, e $>$ 0.05} \\
$[X/Fe]$ &     Fraction (\%)    &    $p$-chance (\%)     &   Fraction (\%)     &          $p$-chance (\%) \\
\hline %\noaling{\smallskip}
$X_{\alpha}$  &  64.0    &       4.0     &       40.0   &  18.5     \\
$X_{\rm Fe}$  &  28.0    &       2.6     &       15.4   &  $<$ 0.1  \\
$X_{\rm vol}$ &  56.0    &      10.4     &       40.0   &  16.8     \\ %% fap = 18.2592%
C/O ratio     &  15.4    &       1.5     &       19.1   &  0.7      \\ % fap = 99.5397%
Mg/Si ratio   &  48.0    &      15.5     &       32.7   &  0.1      \\ %, fap = 99.8652%
\hline
\end{tabular}
\end{scriptsize}
\end{table}

%+++++++++++++++++++++++++++++++++++++++++++++++++++++++++++++++++++
\section{Discussion}\label{seccion_discussion}
%+++++++++++++++++++++++++++++++++++++++++++++++++++++++++++++++++++

  The mechanisms involved in the formation of hot Jupiters are nowadays strongly
  debated (see references in Sect.~\ref{introduccion}).
  In the following we discuss the results from the previous Section
  in the framework of current planet formation and migration models. 
  
%-------------------------------------------------------------------
\subsection{Is our comparison biased?}\label{comp_biased}
%-------------------------------------------------------------------

 The first possibility that we should address is whether or not some stellar
 property is affecting the metal content of any of our samples.
 That is the reason we performed the analysis in Section~\ref{biases} where we found a tendency
%%%As shown in Section~\ref{biases} a tendency 
 of hot PHs to be slightly
 fainter and younger than cool PHs  in our sample.

 \cite{2009ApJ...698L...1H} suggested that the observed correlation between the presence of
 gas-giant planets and high stellar metallicity might be related
 to a possible  galactic inner disc origin of  planet hosts. %these stars. 
 Following this line of reasoning, hints of a correlation between
 the T$_{\rm C}$ slopes (i.e., the linear trends between the
 abundances and the elemental condensation temperature) and
 the stellar age have been noted in the literature
 \citep{2014A&A...564L..15A,2015A&A...579A..20M,2015A&A...579A..52N,2015arXiv151101012S}.
 In a recent study, \cite{2016A&A...588A..98M} found Galactic
 radial mixing to be the only suitable scenario to explain the
 observed T$_{\rm C}$ trends in a large sample of evolved
 (subgiant and red giant) stars. 
 \cite{2014A&A...564L..15A,2016A&A...592A..87A} %and {\bf articulo 2016}
 use the stellar R$_{\rm mean}$  as a proxy of the stellar birthplace
 finding a weak hint that the T$_{\rm C}$ trend depends on R$_{\rm mean}$,
 although the authors highlight the complexity of the dependency. 

 Radial mixing is a secular process and older stars migrate further. 
 Old stars
 might come from a region with significantly different abundances.
 As we have seen there is a tendency of hot PHs to be  younger than
 cool PHs. It could be the case that cool PHs show a wider range of
 metallicities than hot PHs just because they are older.
 This might be in contradiction with the results
 from the  R$_{\rm mean}$ distributions analysis.
 Hot PHs have slightly larger R$_{\rm mean}$ values and stars
 at larger Galactocentric distances (from the outer disc) are expected
 to show lower metallicities \citep[e.g.][Fig. 5]{2008A&A...490..613L}.
 However, the differences in R$_{\rm mean}$ between hot and cool PHs do
 not seem to be significant. 

 Therefore, we conclude that
 we cannot completely rule out the possibility that differences in age and R$_{\rm mean}$
 are affecting the comparison of abundances between both samples.

 The fact that hot PHs are slightly fainter than cool PHs should not,
 to our knowledge, alter the metal content of the stars. %%% (a reference?). 
 Nevertheless, we have explored the $V$ magnitude versus [Fe/H] diagram
 finding no obvious trend.

\begin{figure}[htb]
\centering
\includegraphics[angle=270,scale=0.45]{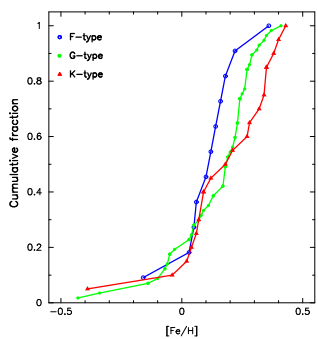}
\caption{
[Fe/H] cumulative frequencies for F-, G-, and K-type stars.
}
\label{fgk_diagram}
\end{figure}

 We finally explore whether the larger fraction of F-type stars in the
 hot PHs (24\%) sample with respect to the cool PHs sample (7\%) might
 affect our results. 
 Although we do not know of any physical reason why F, G, and K stars
 might have different abundance distributions we should carefully check that there are no (unknown)
 systematic effects hidden in the abundance computations that may bias our
 results.
 Figure~\ref{fgk_diagram} shows the metallicity distribution for the F-, G-, and K-type 
 stars. Although no statistically significant differences are found, %%% although
 a tendency of F-type stars to show lower metallicities (at values of [Fe/H] > 0.1 dex) than 
 G- and K-type stars can be seen. Results from a K-S test provide
 $D$ = 0.38, $p$-value = 0.11, n$_{\rm eff}$ = 9.22 when comparing the F and G subsamples.
 A comparison of the K and G subsamples gives $D$ = 0.26, $p$-value = 0.22,
 n$_{\rm eff}$ = 14.81.
 This shows that if different from G, and K-type stars, the metallicity distribution
 of F-type stars would be biased towards lower values at high metallicities.  
 It is therefore unlikely that the larger fraction of F-type stars in the
 hot PHs might explain their higher metallicities when compared with
 the cool PH sample.

%%------------------------------------------------------
%  Figure 9: [Fe/H] cumulative frequencies for F, G, and K-type stars.
%%------------------------------------------------------

\subsection{Metallicity trends and the formation of hot Jupiters}
%-------------------------------------------------------------------

  Until very recently, the existence of hot Jupiters  involved migration
  towards the  star after the planet is formed in metal-rich discs %…….beyond ….(e.g. Pollack et al. 1996)
  at large distances
  from the star, beyond the ice line  ($a$ $\gtrsim$ 1-3 au) where solid material is abundant
  \citep[e.g.][]{1996Icar..124...62P}. On the other hand, the {\it in situ} formation of hot Jupiters has been largely dismissed % discarded
 mainly due to the lack of sufficient condensate solids in the inner 
 regions of the protoplanetary disk \citep{1996Natur.380..606L}. %%%%Rafikov 2006?.
 However, several recent works have pointed out that this assumption
 is mainly based on models of the solar nebula that might not apply
 in general \citep[e.g.][]{2000Icar..143....2B,2016ApJ...817L..17B,2016ApJ...829..114B},
 revisiting whether or not hot Jupiters should necessarily form at large
 distances from the parent star.
  
 To the very best of our knowledge, the exact dependency of the migration mechanisms proposed
 so far (disc or high-eccentricity) on the stellar metallicity is still poorly understood.
 Disc migration might be expected to show some degree of dependence on the stellar metallicity
 \citep[e.g.][]{2016ApJ...823..162L},
 which can be considered a indicator of the disc's opacity 
 (if as usually assumed no alteration of the disc's metal content has occurred).

 The abundance analysis performed in this work might help us in our understanding
 of where hot Jupiters form. Until a better understanding of whether migration can
 alter the metal content of the host stars is achieved, we find it reasonable to assume
 that if  hot Jupiters were formed at large distances from the star and then
 migrate, cool and hot Jupiters should show similar chemical properties.

% \textcolor{blue}{\bf references???, de verdad nadie ha mirado como depende
% la migracion de la metalicidad estelar???} 

% the properties of cool and hot Jupiters %(host stars?)
% should
% be the same
% since none of the  migration mechanisms  proposed so far  (disc or high-eccentricity) have shown
% a direct dependency on stellar metallicity. 
 We find that  more massive planets tend to have larger
 periods in more eccentric orbits. In agreement with previous
 works \citep{2013ApJ...767L..24D,2013A&A...560A..51A} %we note that
 these stars show
 a wider range of metallicities. In other words, hot and cool Jupiters seem to
 constitute two different populations with different properties. 

 As shown in the previous Section there seems to be a ``deficit'' of cool Jupiters in the metallicity
 range +0.00/+0.20 dex. Furthermore, there are no hot Jupiters with
 metallicities below $\sim$ -0.10 dex.
 It should be noted that if both cool and hot Jupiters form at large distances from the star,
 a metallicity dependent migration mechanism should be able to explain
 this difference. %However, none of the known migration mechanisms is known to show a dependence on (or
% expected to modify) the stellar metallicity (a reference is missing).

\begin{figure}[!htb]
\centering
\includegraphics[angle=270,scale=0.45]{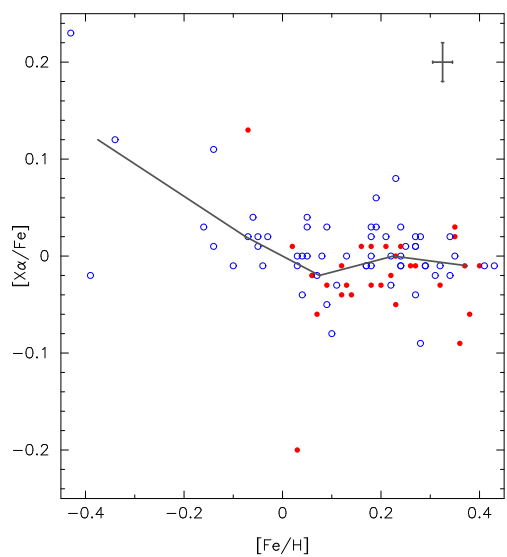}
\caption{
[X$_{\alpha}$/Fe] vs. [Fe/H] plane. Hot PHs are shown by filled red circles while
cool PHs are plotted with open blue circles. The continuous line shows the
mean distribution. 
 Median uncertainties in the derived [X$_{\alpha}$/Fe] and [Fe/H] values
are also shown.}
\label{alpha_vs_metal}
\end{figure}

 The data also shows a tendency of cool PHs to show larger values
 of [X$_{\alpha}$/Fe].  %%%%%% and [X$_{\rm vol}$/Fe].
% The lower volatile abundances in hot-PHs might be explained if these
% elements are preferentially accreted by the planetary core at close distances
% from the star once the forming planet reaches the critical mass and the gas accretion begins. 
% Since these elements are not abundant in the inner regions of the protoplanetary
% disc (where the temperature is very high), once accreted by the forming planet,
% few volatile material remain in the disc to be accreted by the star. 
% If this hypothesis were correct, stars hosting close-in low-mass planets 
% (where no gas accretion has been occurred by the forming planet) do not should show a preference
% for low-volatile abundances. %}
% Nevertheless, results regarding volatile elements should always be
% taken with caution as accurate abundances for these elements are % more
% difficult to derive. 
 The fact that cool PHs show slightly lower metallicity but larger $\alpha$ abundances
 than hot PHs might also provide important clues regarding planet formation. 
  It has been argued that in order to form a sufficiently massive core, the quantity that
  should be considered is the surface density of all condensible elements beyond
  the ice line  \citep{2012A&A...541A..97M}, especially the $\alpha$ elements O, Si, and
  Mg \citep[][]{2006ApJ...643..484R,2009MNRAS.399L.103G}. In particular, it should be
  noted that Mg and Si 
  have condensation temperatures very similar to iron \citep{2003ApJ...591.1220L}.
%Alpha elements have condensation temperatures similar to iron \citep{2003ApJ...591.1220L}
% and have been shown to play an important role in the composition of dust in planet forming regions
% \citep[e.g.][]{2009MNRAS.399L.103G}.
 It is therefore likely that cool PHs might `compensate' their lower metallicity
 content with other contributors to allow planetary formation. %%% (search more references). 
 Along these lines, \cite{2012A&A...543A..89A} found that most of the planet-host stars with low iron content are
 enhanced by $\alpha$ elements. %We think that our results do not contradict
 Figure~\ref{alpha_vs_metal} shows the [X$_{\alpha}$/Fe] versus [Fe/H] plane for our sample.
 While most of our planet hosts are in the metallicity range
 between -0.1 and +0.4 dex where the tendency [X$_{\alpha}$/Fe] versus [Fe/H]
 seems to be flat, a tendency of higher [X$_{\alpha}$/Fe] as we move towards
 lower metallicities seems to be present in our data in agreement
 with the results of \cite{2012A&A...543A..89A}.

%%------------------------------------------------------
%  Figure 10: alpha versus metal plane 
%%------------------------------------------------------

%-------------------------------------------------------------------
\section{Conclusions}\label{conclusions}
%-------------------------------------------------------------------

 In this work a detailed chemical analysis of a large sample of
 gas-giant planet hosts is presented. The sample has been divided
 into stars with cool distant planets and stars with hot Jupiters.
 Before comparing the two subsamples, a detailed analysis of their stellar
 properties was performed to control any possible bias affecting our results.

 The main results of this work can be summarised as follows:

 \begin{itemize}

 \item Hot PHs show higher metallicities than cool PHs
  in the metallicity range between +0.00 and +0.20 dex, and no hot Jupiters
   are found orbiting stars with very low metallicities.

 \item Hot PHs tend to show lower $\alpha$ abundances than cool PHs.
 
 \item Planetary masses of hot Jupiters are typically
  within m$_{\rm p}\sin i$ $\sim$ 1 M$_{\rm Jup}$ and eccentricities do no
  exceed 0.1.

 \item Cool PHs have planetary masses significantly larger than 1 M$_{\rm Jup}$
  and a wider range of eccentricities (between 0.05 and 1).

 \end{itemize}

 We caution that differences in age might be affecting the comparison
 of abundances between both samples, as we find that cool PHs might
 be older than hot PHs. Furthermore, the large fraction of
 F stars in the hot PHs sample should also be considered.

% Our results show that hot-PHs show higher metallicities than cool-PHs
% in the metallicity range between +0.00 and +0.20 dex, and no hot Jupiters
% are found orbiting stars with very low metallicities. %%% lower than -0.10 dex. 
% Hot-PHs tend to show lower $\alpha$ and volatile elements abundances than cool-PHs. 
% Planetary masses of hot Jupiters are typically
% within m$_{\rm p}\sin i$ $\sim$ 1 M$_{\rm Jup}$ and eccentricities do no
% exceed 0.1.
% On the other hand, stars harbouring cool-PHs show a wider range of
% metallicities. Cool-PHs tend to have slightly larger $\alpha$ and volatile elements
% abundances. Planetary masses are significantly larger than 1 M$_{\rm Jup}$
% and the eccentricities can vary between 0.05 and 1. We also find that
% cool-PHs might be older than hot-PHs. 

  Our  results challenge the traditional view that hot Jupiters form
 at large distances from the stars and then migrate. 
 While migration mechanisms might alter some planetary properties, such as,
 for example, the planetary eccentricity, it is unlikely that they
 change the abundance content of the host star.
% Of particular interest is the slightly lower volatile elements
% abundance found in hot Jupiters stars.
% A possibility is that they are preferentially accreted by
% the forming planet but, with the data at hand, this is only an
% hypothesis that should be demonstrated. %}
 The data also show that other elements besides iron, such as 
 Mg, Si, or Ti , might play a role in planet formation by compensating
 the slightly lower metallicity values of cool PHs.

 The detailed chemical analysis of samples complementary to the one 
 analysed here, including different
 planetary types and architectures and different kinds of planet hosts,
 will help us to expand, confirm, or reject  the various trends discussed
 in this work as well as to achieve a comprehensive view of  the planetary formation
 processes. 

%+++++++++++++++++++++++++++++++++++++++++++++++++++++++++++++++++++
\begin{acknowledgements}
%+++++++++++++++++++++++++++++++++++++++++++++++++++++++++++++++++++

      J. M. acknowledges support from the Italian Ministry of Education,
      University, and Research  through the
      \emph{PREMIALE WOW 2013} research project under grant
      \emph{Ricerca di pianeti intorno a stelle di piccola massa}.
      Additional support from the \emph{Ariel ASI-INAF agreement N. 2015-038-R.0}
      is also acknowledged. 
      E. V. and C. E. acknowledge support from the \emph{On the rocks} project
      funded by the Spanish Ministerio de Econom\'ia y Competitividad
      under grant \emph{AYA2014-55840-P}.

\end{acknowledgements}

%<<<<<<<<<<<<<<<<<<<<<<<<<<<<<<<<<<<<<<<<<<<<<<<<<<<<<<<<<<<<<<<<<<<<
%                      Bibliografia
%>>>>>>>>>>>>>>>>>>>>>>>>>>>>>>>>>>>>>>>>>>>>>>>>>>>>>>>>>>>>>>>>>>>>

\bibliographystyle{aa}
\bibliography{hotcool.bib}

\begin{thebibliography}{140}
\expandafter\ifx\csname natexlab\endcsname\relax\def\natexlab#1{#1}\fi

\bibitem[{{Adibekyan} {et~al.}(2016){Adibekyan}, {Delgado-Mena}, {Figueira},
  {Sousa}, {Santos}, {Gonz{\'a}lez Hern{\'a}ndez}, {Minchev}, {Faria},
  {Israelian}, {Harutyunyan}, {Su{\'a}rez-Andr{\'e}s}, \&
  {Hakobyan}}]{2016A&A...592A..87A}
{Adibekyan}, V., {Delgado-Mena}, E., {Figueira}, P., {et~al.} 2016, \aap, 592,
  A87

\bibitem[{{Adibekyan} {et~al.}(2013){Adibekyan}, {Figueira}, {Santos},
  {Mortier}, {Mordasini}, {Delgado Mena}, {Sousa}, {Correia}, {Israelian}, \&
  {Oshagh}}]{2013A&A...560A..51A}
{Adibekyan}, V.~Z., {Figueira}, P., {Santos}, N.~C., {et~al.} 2013, \aap, 560,
  A51

\bibitem[{{Adibekyan} {et~al.}(2014){Adibekyan}, {Gonz{\'a}lez Hern{\'a}ndez},
  {Delgado Mena}, {Sousa}, {Santos}, {Israelian}, {Figueira}, \& {Bertran de
  Lis}}]{2014A&A...564L..15A}
{Adibekyan}, V.~Z., {Gonz{\'a}lez Hern{\'a}ndez}, J.~I., {Delgado Mena}, E.,
  {et~al.} 2014, \aap, 564, L15

\bibitem[{{Adibekyan} {et~al.}(2012){Adibekyan}, {Santos}, {Sousa},
  {Israelian}, {Delgado Mena}, {Gonz{\'a}lez Hern{\'a}ndez}, {Mayor}, {Lovis},
  \& {Udry}}]{2012A&A...543A..89A}
{Adibekyan}, V.~Z., {Santos}, N.~C., {Sousa}, S.~G., {et~al.} 2012, \aap, 543,
  A89

\bibitem[{{Ali-Dib} {et~al.}(2017){Ali-Dib}, {Johansen}, \&
  {Huang}}]{2017MNRAS.469.5016A}
{Ali-Dib}, M., {Johansen}, A., \& {Huang}, C.~X. 2017, \mnras, 469, 5016

\bibitem[{{Alibert} {et~al.}(2005){Alibert}, {Mordasini}, {Benz}, \&
  {Winisdoerffer}}]{2005A&A...434..343A}
{Alibert}, Y., {Mordasini}, C., {Benz}, W., \& {Winisdoerffer}, C. 2005, \aap,
  434, 343

\bibitem[{{Allende Prieto} {et~al.}(2004){Allende Prieto}, {Barklem},
  {Lambert}, \& {Cunha}}]{2004A&A...420..183A}
{Allende Prieto}, C., {Barklem}, P.~S., {Lambert}, D.~L., \& {Cunha}, K. 2004,
  \aap, 420, 183

\bibitem[{{Ammler-von Eiff} {et~al.}(2009){Ammler-von Eiff}, {Santos}, {Sousa},
  {Fernandes}, {Guillot}, {Israelian}, {Mayor}, \&
  {Melo}}]{2009A&A...507..523A}
{Ammler-von Eiff}, M., {Santos}, N.~C., {Sousa}, S.~G., {et~al.} 2009, \aap,
  507, 523

\bibitem[{{Bashi} {et~al.}(2017){Bashi}, {Helled}, {Zucker}, \&
  {Mordasini}}]{2017arXiv170107654B}
{Bashi}, D., {Helled}, R., {Zucker}, S., \& {Mordasini}, C. 2017, \aap, 604,
  A83

\bibitem[{{Batygin} {et~al.}(2016){Batygin}, {Bodenheimer}, \&
  {Laughlin}}]{2016ApJ...829..114B}
{Batygin}, K., {Bodenheimer}, P.~H., \& {Laughlin}, G.~P. 2016, \apj, 829, 114

\bibitem[{{Beaug{\'e}} \& {Nesvorn{\'y}}(2012)}]{2012ApJ...751..119B}
{Beaug{\'e}}, C. \& {Nesvorn{\'y}}, D. 2012, \apj, 751, 119

\bibitem[{{Beaug{\'e}} \& {Nesvorn{\'y}}(2013)}]{2013ApJ...763...12B}
{Beaug{\'e}}, C. \& {Nesvorn{\'y}}, D. 2013, \apj, 763, 12

\bibitem[{{Bensby} {et~al.}(2003){Bensby}, {Feltzing}, \&
  {Lundstr{\"o}m}}]{2003A&A...410..527B}
{Bensby}, T., {Feltzing}, S., \& {Lundstr{\"o}m}, I. 2003, \aap, 410, 527

\bibitem[{{Bensby} {et~al.}(2005){Bensby}, {Feltzing}, {Lundstr{\"o}m}, \&
  {Ilyin}}]{2005A&A...433..185B}
{Bensby}, T., {Feltzing}, S., {Lundstr{\"o}m}, I., \& {Ilyin}, I. 2005, \aap,
  433, 185

\bibitem[{{Bodenheimer} {et~al.}(2000){Bodenheimer}, {Hubickyj}, \&
  {Lissauer}}]{2000Icar..143....2B}
{Bodenheimer}, P., {Hubickyj}, O., \& {Lissauer}, J.~J. 2000, \icarus, 143, 2

\bibitem[{{Boley} {et~al.}(2016){Boley}, {Granados Contreras}, \&
  {Gladman}}]{2016ApJ...817L..17B}
{Boley}, A.~C., {Granados Contreras}, A.~P., \& {Gladman}, B. 2016, \apjl, 817,
  L17

\bibitem[{{Boley} {et~al.}(2012){Boley}, {Payne}, \&
  {Ford}}]{2012ApJ...754...57B}
{Boley}, A.~C., {Payne}, M.~J., \& {Ford}, E.~B. 2012, \apj, 754, 57

\bibitem[{{Bond} {et~al.}(2010){Bond}, {O'Brien}, \&
  {Lauretta}}]{2010ApJ...715.1050B}
{Bond}, J.~C., {O'Brien}, D.~P., \& {Lauretta}, D.~S. 2010, \apj, 715, 1050

\bibitem[{{Boss}(1995)}]{1995Sci...267..360B}
{Boss}, A.~P. 1995, Science, 267, 360

\bibitem[{{Boss}(1997)}]{1997Sci...276.1836B}
{Boss}, A.~P. 1997, Science, 276, 1836

\bibitem[{{Boss}(2002)}]{2002ApJ...567L.149B}
{Boss}, A.~P. 2002, \apjl, 567, L149

\bibitem[{{Boss}(2006)}]{2006ApJ...643..501B}
{Boss}, A.~P. 2006, \apj, 643, 501

\bibitem[{{Bouchy} \& {Sophie Team}(2006)}]{2006tafp.conf..319B}
{Bouchy}, F. \& {Sophie Team}. 2006, in Tenth Anniversary of 51 Peg-b: Status
  of and prospects for hot Jupiter studies, ed. L.~{Arnold}, F.~{Bouchy}, \&
  C.~{Moutou}, 319--325

\bibitem[{{Bressan} {et~al.}(2012){Bressan}, {Marigo}, {Girardi}, {Salasnich},
  {Dal Cero}, {Rubele}, \& {Nanni}}]{2012MNRAS.427..127B}
{Bressan}, A., {Marigo}, P., {Girardi}, L., {et~al.} 2012, \mnras, 427, 127

\bibitem[{{Bromley} \& {Kenyon}(2011)}]{2011ApJ...735...29B}
{Bromley}, B.~C. \& {Kenyon}, S.~J. 2011, \apj, 735, 29

\bibitem[{{Bryan} {et~al.}(2016){Bryan}, {Knutson}, {Howard}, {Ngo}, {Batygin},
  {Crepp}, {Fulton}, {Hinkley}, {Isaacson}, {Johnson}, {Marcy}, \&
  {Wright}}]{2016ApJ...821...89B}
{Bryan}, M.~L., {Knutson}, H.~A., {Howard}, A.~W., {et~al.} 2016, \apj, 821, 89

\bibitem[{{Buchhave} \& {Latham}(2015)}]{2015ApJ...808..187B}
{Buchhave}, L.~A. \& {Latham}, D.~W. 2015, \apj, 808, 187

\bibitem[{{Buchhave} {et~al.}(2012){Buchhave}, {Latham}, {Johansen},
  {Bizzarro}, {Torres}, {Rowe}, {Batalha}, {Borucki}, {Brugamyer}, {Caldwell},
  {Bryson}, {Ciardi}, {Cochran}, {Endl}, {Esquerdo}, {Ford}, {Geary},
  {Gilliland}, {Hansen}, {Isaacson}, {Laird}, {Lucas}, {Marcy}, {Morse},
  {Robertson}, {Shporer}, {Stefanik}, {Still}, \&
  {Quinn}}]{2012Natur.486..375B}
{Buchhave}, L.~A., {Latham}, D.~W., {Johansen}, A., {et~al.} 2012, \nat, 486,
  375

\bibitem[{{Butler} {et~al.}(1997){Butler}, {Marcy}, {Williams}, {Hauser}, \&
  {Shirts}}]{1997ApJ...474L.115B}
{Butler}, R.~P., {Marcy}, G.~W., {Williams}, E., {Hauser}, H., \& {Shirts}, P.
  1997, \apjl, 474, L115

\bibitem[{{Casagrande} {et~al.}(2011){Casagrande}, {Sch{\"o}nrich}, {Asplund},
  {Cassisi}, {Ram{\'{\i}}rez}, {Mel{\'e}ndez}, {Bensby}, \&
  {Feltzing}}]{2011A&A...530A.138C}
{Casagrande}, L., {Sch{\"o}nrich}, R., {Asplund}, M., {et~al.} 2011, \aap, 530,
  A138

\bibitem[{{Cassan} {et~al.}(2012){Cassan}, {Kubas}, {Beaulieu}, {Dominik},
  {Horne}, {Greenhill}, {Wambsganss}, {Menzies}, {Williams}, {J{\o}rgensen},
  {Udalski}, {Bennett}, {Albrow}, {Batista}, {Brillant}, {Caldwell}, {Cole},
  {Coutures}, {Cook}, {Dieters}, {Prester}, {Donatowicz}, {Fouqu{\'e}}, {Hill},
  {Kains}, {Kane}, {Marquette}, {Martin}, {Pollard}, {Sahu}, {Vinter},
  {Warren}, {Watson}, {Zub}, {Sumi}, {Szyma{\'n}ski}, {Kubiak}, {Poleski},
  {Soszynski}, {Ulaczyk}, {Pietrzy{\'n}ski}, \&
  {Wyrzykowski}}]{2012Natur.481..167C}
{Cassan}, A., {Kubas}, D., {Beaulieu}, J.-P., {et~al.} 2012, \nat, 481, 167

\bibitem[{{Chatterjee} {et~al.}(2008){Chatterjee}, {Ford}, {Matsumura}, \&
  {Rasio}}]{2008ApJ...686..580C}
{Chatterjee}, S., {Ford}, E.~B., {Matsumura}, S., \& {Rasio}, F.~A. 2008, \apj,
  686, 580

\bibitem[{{Chiang} \& {Laughlin}(2013)}]{2013MNRAS.431.3444C}
{Chiang}, E. \& {Laughlin}, G. 2013, \mnras, 431, 3444

\bibitem[{{Cumming} {et~al.}(2008){Cumming}, {Butler}, {Marcy}, {Vogt},
  {Wright}, \& {Fischer}}]{2008PASP..120..531C}
{Cumming}, A., {Butler}, R.~P., {Marcy}, G.~W., {et~al.} 2008, \pasp, 120, 531

\bibitem[{{Currie}(2009)}]{2009ApJ...694L.171C}
{Currie}, T. 2009, \apjl, 694, L171

\bibitem[{{da Silva} {et~al.}(2006){da Silva}, {Girardi}, {Pasquini},
  {Setiawan}, {von der L{\"u}he}, {de Medeiros}, {Hatzes}, {D{\"o}llinger}, \&
  {Weiss}}]{2006A&A...458..609D}
{da Silva}, L., {Girardi}, L., {Pasquini}, L., {et~al.} 2006, \aap, 458, 609

\bibitem[{{Dawson} \& {Murray-Clay}(2013)}]{2013ApJ...767L..24D}
{Dawson}, R.~I. \& {Murray-Clay}, R.~A. 2013, \apjl, 767, L24

\bibitem[{{Dodson-Robinson} {et~al.}(2009){Dodson-Robinson}, {Veras}, {Ford},
  \& {Beichman}}]{2009ApJ...707...79D}
{Dodson-Robinson}, S.~E., {Veras}, D., {Ford}, E.~B., \& {Beichman}, C.~A.
  2009, \apj, 707, 79

\bibitem[{{Dong} {et~al.}(2017){Dong}, {Xie}, {Zhou}, {Zheng}, \&
  {Luo}}]{2017arXiv170607807D}
{Dong}, S., {Xie}, J.-W., {Zhou}, J.-L., {Zheng}, Z., \& {Luo}, A. 2017, ArXiv
  e-prints [\eprint[arXiv]{1706.07807}]

\bibitem[{{Dorn} {et~al.}(2015){Dorn}, {Khan}, {Heng}, {Connolly}, {Alibert},
  {Benz}, \& {Tackley}}]{2015A&A...577A..83D}
{Dorn}, C., {Khan}, A., {Heng}, K., {et~al.} 2015, \aap, 577, A83

\bibitem[{{ESA}(1997)}]{1997ESASP1200.....E}
{ESA}, ed. 1997, ESA Special Publication, Vol. 1200, {The HIPPARCOS and TYCHO
  catalogues. Astrometric and photometric star catalogues derived from the ESA
  HIPPARCOS Space Astrometry Mission}

\bibitem[{{Fabrycky} \& {Tremaine}(2007)}]{2007ApJ...669.1298F}
{Fabrycky}, D. \& {Tremaine}, S. 2007, \apj, 669, 1298

\bibitem[{{Fischer} \& {Valenti}(2005)}]{2005ApJ...622.1102F}
{Fischer}, D.~A. \& {Valenti}, J. 2005, \apj, 622, 1102

\bibitem[{{Ford} \& {Rasio}(2006)}]{2006ApJ...638L..45F}
{Ford}, E.~B. \& {Rasio}, F.~A. 2006, \apjl, 638, L45

\bibitem[{{Ford} \& {Rasio}(2008)}]{2008ApJ...686..621F}
{Ford}, E.~B. \& {Rasio}, F.~A. 2008, \apj, 686, 621

\bibitem[{{Fressin} {et~al.}(2013){Fressin}, {Torres}, {Charbonneau}, {Bryson},
  {Christiansen}, {Dressing}, {Jenkins}, {Walkowicz}, \&
  {Batalha}}]{2013ApJ...766...81F}
{Fressin}, F., {Torres}, G., {Charbonneau}, D., {et~al.} 2013, \apj, 766, 81

\bibitem[{{Ghezzi} {et~al.}(2010{\natexlab{a}}){Ghezzi}, {Cunha}, {Schuler}, \&
  {Smith}}]{2010ApJ...725..721G}
{Ghezzi}, L., {Cunha}, K., {Schuler}, S.~C., \& {Smith}, V.~V.
  2010{\natexlab{a}}, \apj, 725, 721

\bibitem[{{Ghezzi} {et~al.}(2010{\natexlab{b}}){Ghezzi}, {Cunha}, {Smith}, {de
  Ara{\'u}jo}, {Schuler}, \& {de la Reza}}]{2010ApJ...720.1290G}
{Ghezzi}, L., {Cunha}, K., {Smith}, V.~V., {et~al.} 2010{\natexlab{b}}, \apj,
  720, 1290

\bibitem[{{Goldreich} \& {Tremaine}(1980)}]{1980ApJ...241..425G}
{Goldreich}, P. \& {Tremaine}, S. 1980, \apj, 241, 425

\bibitem[{{Gonzalez}(2009)}]{2009MNRAS.399L.103G}
{Gonzalez}, G. 2009, \mnras, 399, L103

\bibitem[{{Gonz{\'a}lez Hern{\'a}ndez} {et~al.}(2013){Gonz{\'a}lez
  Hern{\'a}ndez}, {Delgado-Mena}, {Sousa}, {Israelian}, {Santos}, {Adibekyan},
  \& {Udry}}]{2013A&A...552A...6G}
{Gonz{\'a}lez Hern{\'a}ndez}, J.~I., {Delgado-Mena}, E., {Sousa}, S.~G.,
  {et~al.} 2013, \aap, 552, A6

\bibitem[{{Gould} {et~al.}(2006){Gould}, {Dorsher}, {Gaudi}, \&
  {Udalski}}]{2006AcA....56....1G}
{Gould}, A., {Dorsher}, S., {Gaudi}, B.~S., \& {Udalski}, A. 2006, \actaa, 56,
  1

\bibitem[{{Hamers} {et~al.}(2016){Hamers}, {Perets}, \& {Portegies
  Zwart}}]{2016MNRAS.455.3180H}
{Hamers}, A.~S., {Perets}, H.~B., \& {Portegies Zwart}, S.~F. 2016, \mnras,
  455, 3180

\bibitem[{{Haywood}(2009)}]{2009ApJ...698L...1H}
{Haywood}, M. 2009, \apjl, 698, L1

\bibitem[{{Hekker} \& {Mel{\'e}ndez}(2007)}]{2007A&A...475.1003H}
{Hekker}, S. \& {Mel{\'e}ndez}, J. 2007, \aap, 475, 1003

\bibitem[{{H{\o}g} {et~al.}(2000){H{\o}g}, {Fabricius}, {Makarov}, {Bastian},
  {Schwekendiek}, {Wicenec}, {Urban}, {Corbin}, \&
  {Wycoff}}]{2000A&A...357..367H}
{H{\o}g}, E., {Fabricius}, C., {Makarov}, V.~V., {et~al.} 2000, \aap, 357, 367

\bibitem[{{Howard} {et~al.}(2012){Howard}, {Marcy}, {Bryson}, {Jenkins},
  {Rowe}, {Batalha}, {Borucki}, {Koch}, {Dunham}, {Gautier}, {Van Cleve},
  {Cochran}, {Latham}, {Lissauer}, {Torres}, {Brown}, {Gilliland}, {Buchhave},
  {Caldwell}, {Christensen-Dalsgaard}, {Ciardi}, {Fressin}, {Haas}, {Howell},
  {Kjeldsen}, {Seager}, {Rogers}, {Sasselov}, {Steffen}, {Basri},
  {Charbonneau}, {Christiansen}, {Clarke}, {Dupree}, {Fabrycky}, {Fischer},
  {Ford}, {Fortney}, {Tarter}, {Girouard}, {Holman}, {Johnson}, {Klaus},
  {Machalek}, {Moorhead}, {Morehead}, {Ragozzine}, {Tenenbaum}, {Twicken},
  {Quinn}, {Isaacson}, {Shporer}, {Lucas}, {Walkowicz}, {Welsh}, {Boss},
  {Devore}, {Gould}, {Smith}, {Morris}, {Prsa}, {Morton}, {Still}, {Thompson},
  {Mullally}, {Endl}, \& {MacQueen}}]{2012ApJS..201...15H}
{Howard}, A.~W., {Marcy}, G.~W., {Bryson}, S.~T., {et~al.} 2012, \apjs, 201, 15

\bibitem[{{Huang} {et~al.}(2016){Huang}, {Wu}, \&
  {Triaud}}]{2016ApJ...825...98H}
{Huang}, C., {Wu}, Y., \& {Triaud}, A.~H.~M.~J. 2016, \apj, 825, 98

\bibitem[{{Ida} \& {Lin}(2008)}]{2008ApJ...673..487I}
{Ida}, S. \& {Lin}, D.~N.~C. 2008, \apj, 673, 487

\bibitem[{{Jenkins} {et~al.}(2017){Jenkins}, {Jones}, {Tuomi}, {D{\'{\i}}az},
  {Cordero}, {Aguayo}, {Pantoja}, {Arriagada}, {Mahu}, {Brahm}, {Rojo}, {Soto},
  {Ivanyuk}, {Becerra Yoma}, {Day-Jones}, {Ruiz}, {Pavlenko}, {Barnes},
  {Murgas}, {Pinfield}, {Jones}, {L{\'o}pez-Morales}, {Shectman}, {Butler}, \&
  {Minniti}}]{2017MNRAS.466..443J}
{Jenkins}, J.~S., {Jones}, H.~R.~A., {Tuomi}, M., {et~al.} 2017, \mnras, 466,
  443

\bibitem[{{Jofr{\'e}} {et~al.}(2015){Jofr{\'e}}, {Petrucci}, {Saffe}, {Saker},
  {de la Villarmois}, {Chavero}, {G{\'o}mez}, \& {Mauas}}]{2015A&A...574A..50J}
{Jofr{\'e}}, E., {Petrucci}, R., {Saffe}, C., {et~al.} 2015, \aap, 574, A50

\bibitem[{{Johansen} \& {Lacerda}(2010)}]{2010MNRAS.404..475J}
{Johansen}, A. \& {Lacerda}, P. 2010, \mnras, 404, 475

\bibitem[{{Johansen} \& {Lambrechts}(2017)}]{2017AREPS..45..359J}
{Johansen}, A. \& {Lambrechts}, M. 2017, Annual Review of Earth and Planetary
  Sciences, 45, 359

\bibitem[{{Jones} {et~al.}(2003){Jones}, {Butler}, {Tinney}, {Marcy}, {Penny},
  {McCarthy}, \& {Carter}}]{2003MNRAS.341..948J}
{Jones}, H.~R.~A., {Butler}, R.~P., {Tinney}, C.~G., {et~al.} 2003, \mnras,
  341, 948

\bibitem[{{Juri{\'c}} \& {Tremaine}(2008)}]{2008ApJ...686..603J}
{Juri{\'c}}, M. \& {Tremaine}, S. 2008, \apj, 686, 603

\bibitem[{{Kaufer} {et~al.}(1999){Kaufer}, {Stahl}, {Tubbesing},
  {N{\o}rregaard}, {Avila}, {Francois}, {Pasquini}, \&
  {Pizzella}}]{1999Msngr..95....8K}
{Kaufer}, A., {Stahl}, O., {Tubbesing}, S., {et~al.} 1999, The Messenger, 95, 8

\bibitem[{{Kiselman}(1993)}]{1993A&A...275..269K}
{Kiselman}, D. 1993, \aap, 275, 269

\bibitem[{{Kiselman}(2001)}]{2001NewAR..45..559K}
{Kiselman}, D. 2001, \nar, 45, 559

\bibitem[{{Kley} \& {Nelson}(2012)}]{2012ARA&A..50..211K}
{Kley}, W. \& {Nelson}, R.~P. 2012, \araa, 50, 211

\bibitem[{{Knutson} {et~al.}(2014){Knutson}, {Fulton}, {Montet}, {Kao}, {Ngo},
  {Howard}, {Crepp}, {Hinkley}, {Bakos}, {Batygin}, {Johnson}, {Morton}, \&
  {Muirhead}}]{2014ApJ...785..126K}
{Knutson}, H.~A., {Fulton}, B.~J., {Montet}, B.~T., {et~al.} 2014, \apj, 785,
  126

\bibitem[{{Kozai}(1962)}]{1962AJ.....67R.579K}
{Kozai}, Y. 1962, \aj, 67, 579

\bibitem[{{Kurucz}(1993)}]{1993KurCD..13.....K}
{Kurucz}, R. 1993, ATLAS9 Stellar Atmosphere Programs and 2 km/s grid.~Kurucz
  CD-ROM No.~13.~ Cambridge, Mass.: Smithsonian Astrophysical Observatory,
  1993., 13

\bibitem[{{Lambrechts} \& {Johansen}(2012)}]{2012A&A...544A..32L}
{Lambrechts}, M. \& {Johansen}, A. 2012, \aap, 544, A32

\bibitem[{{Lambrechts} \& {Johansen}(2014)}]{2014A&A...572A.107L}
{Lambrechts}, M. \& {Johansen}, A. 2014, \aap, 572, A107

\bibitem[{{Latham} {et~al.}(2011){Latham}, {Rowe}, {Quinn}, {Batalha},
  {Borucki}, {Brown}, {Bryson}, {Buchhave}, {Caldwell}, {Carter},
  {Christiansen}, {Ciardi}, {Cochran}, {Dunham}, {Fabrycky}, {Ford}, {Gautier},
  {Gilliland}, {Holman}, {Howell}, {Ibrahim}, {Isaacson}, {Jenkins}, {Koch},
  {Lissauer}, {Marcy}, {Quintana}, {Ragozzine}, {Sasselov}, {Shporer},
  {Steffen}, {Welsh}, \& {Wohler}}]{2011ApJ...732L..24L}
{Latham}, D.~W., {Rowe}, J.~F., {Quinn}, S.~N., {et~al.} 2011, \apjl, 732, L24

\bibitem[{{Lemasle} {et~al.}(2008){Lemasle}, {Fran{\c c}ois}, {Piersimoni},
  {Pedicelli}, {Bono}, {Laney}, {Primas}, \&
  {Romaniello}}]{2008A&A...490..613L}
{Lemasle}, B., {Fran{\c c}ois}, P., {Piersimoni}, A., {et~al.} 2008, \aap, 490,
  613

\bibitem[{{Lidov}(1962)}]{1962P&SS....9..719L}
{Lidov}, M.~L. 1962, \planss, 9, 719

\bibitem[{{Lin} {et~al.}(1996){Lin}, {Bodenheimer}, \&
  {Richardson}}]{1996Natur.380..606L}
{Lin}, D.~N.~C., {Bodenheimer}, P., \& {Richardson}, D.~C. 1996, \nat, 380, 606

\bibitem[{{Liu} {et~al.}(2016){Liu}, {Zhang}, \& {Lin}}]{2016ApJ...823..162L}
{Liu}, B., {Zhang}, X., \& {Lin}, D.~N.~C. 2016, \apj, 823, 162

\bibitem[{{Lodders}(2003)}]{2003ApJ...591.1220L}
{Lodders}, K. 2003, \apj, 591, 1220

\bibitem[{{Malaroda} {et~al.}(2006){Malaroda}, {Levato}, \&
  {Galliani}}]{2006yCat.3249....0M}
{Malaroda}, S., {Levato}, H., \& {Galliani}, S. 2006, VizieR Online Data
  Catalog, 3249

\bibitem[{{Maldonado} {et~al.}(2015){Maldonado}, {Eiroa}, {Villaver},
  {Montesinos}, \& {Mora}}]{2015A&A...579A..20M}
{Maldonado}, J., {Eiroa}, C., {Villaver}, E., {Montesinos}, B., \& {Mora}, A.
  2015, \aap, 579, A20

\bibitem[{{Maldonado} \& {Villaver}(2016)}]{2016A&A...588A..98M}
{Maldonado}, J. \& {Villaver}, E. 2016, \aap, 588, A98

\bibitem[{{Maldonado} \& {Villaver}(2017)}]{2017A&A...602A..38M}
{Maldonado}, J. \& {Villaver}, E. 2017, \aap, 602, A38

\bibitem[{{Maldonado} {et~al.}(2013){Maldonado}, {Villaver}, \&
  {Eiroa}}]{2013A&A...554A..84M}
{Maldonado}, J., {Villaver}, E., \& {Eiroa}, C. 2013, \aap, 554, A84

\bibitem[{{Mata S{\'a}nchez} {et~al.}(2014){Mata S{\'a}nchez}, {Gonz{\'a}lez
  Hern{\'a}ndez}, {Israelian}, {Santos}, {Sahlmann}, \&
  {Udry}}]{2014A&A...566A..83M}
{Mata S{\'a}nchez}, D., {Gonz{\'a}lez Hern{\'a}ndez}, J.~I., {Israelian}, G.,
  {et~al.} 2014, \aap, 566, A83

\bibitem[{{Matsumura} {et~al.}(2010){Matsumura}, {Peale}, \&
  {Rasio}}]{2010ApJ...725.1995M}
{Matsumura}, S., {Peale}, S.~J., \& {Rasio}, F.~A. 2010, \apj, 725, 1995

\bibitem[{{Mayor} {et~al.}(2011){Mayor}, {Marmier}, {Lovis}, {Udry},
  {S{\'e}gransan}, {Pepe}, {Benz}, {Bertaux}, {Bouchy}, {Dumusque}, {Lo Curto},
  {Mordasini}, {Queloz}, \& {Santos}}]{2011arXiv1109.2497M}
{Mayor}, M., {Marmier}, M., {Lovis}, C., {et~al.} 2011, ArXiv e-prints
  [\eprint[arXiv]{1109.2497}]

\bibitem[{{Mayor} {et~al.}(2003){Mayor}, {Pepe}, {Queloz}, {Bouchy},
  {Rupprecht}, {Lo Curto}, {Avila}, {Benz}, {Bertaux}, {Bonfils}, {Dall},
  {Dekker}, {Delabre}, {Eckert}, {Fleury}, {Gilliotte}, {Gojak}, {Guzman},
  {Kohler}, {Lizon}, {Longinotti}, {Lovis}, {Megevand}, {Pasquini}, {Reyes},
  {Sivan}, {Sosnowska}, {Soto}, {Udry}, {van Kesteren}, {Weber}, \&
  {Weilenmann}}]{2003Msngr.114...20M}
{Mayor}, M., {Pepe}, F., {Queloz}, D., {et~al.} 2003, The Messenger, 114, 20

\bibitem[{{Mayor} \& {Queloz}(1995)}]{1995Natur.378..355M}
{Mayor}, M. \& {Queloz}, D. 1995, \nat, 378, 355

\bibitem[{{Mel{\'e}ndez} {et~al.}(2009){Mel{\'e}ndez}, {Asplund}, {Gustafsson},
  \& {Yong}}]{2009ApJ...704L..66M}
{Mel{\'e}ndez}, J., {Asplund}, M., {Gustafsson}, B., \& {Yong}, D. 2009, \apjl,
  704, L66

\bibitem[{{Montes} {et~al.}(2001){Montes}, {L{\'o}pez-Santiago}, {G{\'a}lvez},
  {Fern{\'a}ndez-Figueroa}, {De Castro}, \& {Cornide}}]{2001MNRAS.328...45M}
{Montes}, D., {L{\'o}pez-Santiago}, J., {G{\'a}lvez}, M.~C., {et~al.} 2001,
  \mnras, 328, 45

\bibitem[{{Mordasini} {et~al.}(2009){Mordasini}, {Alibert}, \&
  {Benz}}]{2009A&A...501.1139M}
{Mordasini}, C., {Alibert}, Y., \& {Benz}, W. 2009, \aap, 501, 1139

\bibitem[{{Mordasini} {et~al.}(2012){Mordasini}, {Alibert}, {Benz}, {Klahr}, \&
  {Henning}}]{2012A&A...541A..97M}
{Mordasini}, C., {Alibert}, Y., {Benz}, W., {Klahr}, H., \& {Henning}, T. 2012,
  \aap, 541, A97

\bibitem[{{Mortier} {et~al.}(2013){Mortier}, {Santos}, {Sousa}, {Adibekyan},
  {Delgado Mena}, {Tsantaki}, {Israelian}, \& {Mayor}}]{2013A&A...557A..70M}
{Mortier}, A., {Santos}, N.~C., {Sousa}, S.~G., {et~al.} 2013, \aap, 557, A70

\bibitem[{{Nagasawa} \& {Ida}(2011)}]{2011ApJ...742...72N}
{Nagasawa}, M. \& {Ida}, S. 2011, \apj, 742, 72

\bibitem[{{Naoz} {et~al.}(2012){Naoz}, {Farr}, \&
  {Rasio}}]{2012ApJ...754L..36N}
{Naoz}, S., {Farr}, W.~M., \& {Rasio}, F.~A. 2012, \apjl, 754, L36

\bibitem[{{Nissen}(2015)}]{2015A&A...579A..52N}
{Nissen}, P.~E. 2015, \aap, 579, A52

\bibitem[{{Ormel} \& {Klahr}(2010)}]{2010A&A...520A..43O}
{Ormel}, C.~W. \& {Klahr}, H.~H. 2010, \aap, 520, A43

\bibitem[{{Papaloizou} \& {Larwood}(2000)}]{2000MNRAS.315..823P}
{Papaloizou}, J.~C.~B. \& {Larwood}, J.~D. 2000, \mnras, 315, 823

\bibitem[{{Pasquini} {et~al.}(2007){Pasquini}, {D{\"o}llinger}, {Weiss},
  {Girardi}, {Chavero}, {Hatzes}, {da Silva}, \&
  {Setiawan}}]{2007A&A...473..979P}
{Pasquini}, L., {D{\"o}llinger}, M.~P., {Weiss}, A., {et~al.} 2007, \aap, 473,
  979

\bibitem[{{Petigura} {et~al.}(2013){Petigura}, {Howard}, \&
  {Marcy}}]{2013PNAS..11019273P}
{Petigura}, E.~A., {Howard}, A.~W., \& {Marcy}, G.~W. 2013, Proceedings of the
  National Academy of Science, 110, 19273

\bibitem[{{Pollack} {et~al.}(1996){Pollack}, {Hubickyj}, {Bodenheimer},
  {Lissauer}, {Podolak}, \& {Greenzweig}}]{1996Icar..124...62P}
{Pollack}, J.~B., {Hubickyj}, O., {Bodenheimer}, P., {et~al.} 1996, \icarus,
  124, 62

\bibitem[{{Rafikov}(2006)}]{2006ApJ...648..666R}
{Rafikov}, R.~R. 2006, \apj, 648, 666

\bibitem[{{Rafikov}(2011)}]{2011ApJ...727...86R}
{Rafikov}, R.~R. 2011, \apj, 727, 86

\bibitem[{{Ram{\'{\i}}rez} {et~al.}(2009){Ram{\'{\i}}rez}, {Mel{\'e}ndez}, \&
  {Asplund}}]{2009A&A...508L..17R}
{Ram{\'{\i}}rez}, I., {Mel{\'e}ndez}, J., \& {Asplund}, M. 2009, \aap, 508, L17

\bibitem[{{Rasio} \& {Ford}(1996)}]{1996Sci...274..954R}
{Rasio}, F.~A. \& {Ford}, E.~B. 1996, Science, 274, 954

\bibitem[{{Reffert} {et~al.}(2015){Reffert}, {Bergmann}, {Quirrenbach},
  {Trifonov}, \& {K{\"u}nstler}}]{2015A&A...574A.116R}
{Reffert}, S., {Bergmann}, C., {Quirrenbach}, A., {Trifonov}, T., \&
  {K{\"u}nstler}, A. 2015, \aap, 574, A116

\bibitem[{{Ribas} \& {Miralda-Escud{\'e}}(2007)}]{2007A&A...464..779R}
{Ribas}, I. \& {Miralda-Escud{\'e}}, J. 2007, \aap, 464, 779

\bibitem[{{Robinson} {et~al.}(2006){Robinson}, {Laughlin}, {Bodenheimer}, \&
  {Fischer}}]{2006ApJ...643..484R}
{Robinson}, S.~E., {Laughlin}, G., {Bodenheimer}, P., \& {Fischer}, D. 2006,
  \apj, 643, 484

\bibitem[{{Sadakane} {et~al.}(2005){Sadakane}, {Ohnishi}, {Ohkubo}, \&
  {Takeda}}]{2005PASJ...57..127S}
{Sadakane}, K., {Ohnishi}, T., {Ohkubo}, M., \& {Takeda}, Y. 2005, \pasj, 57,
  127

\bibitem[{{Santerne} {et~al.}(2016){Santerne}, {Moutou}, {Tsantaki}, {Bouchy},
  {H{\'e}brard}, {Adibekyan}, {Almenara}, {Amard}, {Barros}, {Boisse},
  {Bonomo}, {Bruno}, {Courcol}, {Deleuil}, {Demangeon}, {D{\'{\i}}az},
  {Guillot}, {Havel}, {Montagnier}, {Rajpurohit}, {Rey}, \&
  {Santos}}]{2016A&A...587A..64S}
{Santerne}, A., {Moutou}, C., {Tsantaki}, M., {et~al.} 2016, \aap, 587, A64

\bibitem[{{Santos} {et~al.}(2017){Santos}, {Adibekyan}, {Figueira},
  {Andreasen}, {Barros}, {Delgado-Mena}, {Demangeon}, {Faria}, {Oshagh},
  {Sousa}, {Viana}, \& {Ferreira}}]{2017A&A...603A..30S}
{Santos}, N.~C., {Adibekyan}, V., {Figueira}, P., {et~al.} 2017, \aap, 603, A30

\bibitem[{{Santos} {et~al.}(2003){Santos}, {Israelian}, {Mayor}, {Rebolo}, \&
  {Udry}}]{2003A&A...398..363S}
{Santos}, N.~C., {Israelian}, G., {Mayor}, M., {Rebolo}, R., \& {Udry}, S.
  2003, \aap, 398, 363

\bibitem[{{Schlaufman} \& {Winn}(2016)}]{2016ApJ...825...62S}
{Schlaufman}, K.~C. \& {Winn}, J.~N. 2016, \apj, 825, 62

\bibitem[{{Schneider} {et~al.}(2011){Schneider}, {Dedieu}, {Le Sidaner},
  {Savalle}, \& {Zolotukhin}}]{2011A&A...532A..79S}
{Schneider}, J., {Dedieu}, C., {Le Sidaner}, P., {Savalle}, R., \&
  {Zolotukhin}, I. 2011, \aap, 532, A79

\bibitem[{{Schuler} {et~al.}(2005){Schuler}, {Kim}, {Tinker}, {King}, {Hatzes},
  \& {Guenther}}]{2005ApJ...632L.131S}
{Schuler}, S.~C., {Kim}, J.~H., {Tinker}, Jr., M.~C., {et~al.} 2005, \apjl,
  632, L131

\bibitem[{{Schuler} {et~al.}(2015){Schuler}, {Vaz}, {Katime Santrich}, {Cunha},
  {Smith}, {King}, {Teske}, {Ghezzi}, {Howell}, \&
  {Isaacson}}]{2015ApJ...815....5S}
{Schuler}, S.~C., {Vaz}, Z.~A., {Katime Santrich}, O.~J., {et~al.} 2015, \apj,
  815, 5

\bibitem[{{Sneden}(1973)}]{1973PhDT.......180S}
{Sneden}, C.~A. 1973, PhD thesis, THE UNIVERSITY OF TEXAS AT AUSTIN.

\bibitem[{{Sousa} {et~al.}(2011){Sousa}, {Santos}, {Israelian}, {Mayor}, \&
  {Udry}}]{2011A&A...533A.141S}
{Sousa}, S.~G., {Santos}, N.~C., {Israelian}, G., {Mayor}, M., \& {Udry}, S.
  2011, \aap, 533, A141

\bibitem[{{Sozzetti}(2004)}]{2004MNRAS.354.1194S}
{Sozzetti}, A. 2004, \mnras, 354, 1194

\bibitem[{{Spina} {et~al.}(2016){Spina}, {Mel{\'e}ndez}, \&
  {Ram{\'{\i}}rez}}]{2015arXiv151101012S}
{Spina}, L., {Mel{\'e}ndez}, J., \& {Ram{\'{\i}}rez}, I. 2016, \aap, 585, A152

\bibitem[{{Steffen} {et~al.}(2012){Steffen}, {Ragozzine}, {Fabrycky}, {Carter},
  {Ford}, {Holman}, {Rowe}, {Welsh}, {Borucki}, {Boss}, {Ciardi}, \&
  {Quinn}}]{2012PNAS..109.7982S}
{Steffen}, J.~H., {Ragozzine}, D., {Fabrycky}, D.~C., {et~al.} 2012,
  Proceedings of the National Academy of Science, 109, 7982

\bibitem[{{Takeda}(2003)}]{2003A&A...402..343T}
{Takeda}, Y. 2003, \aap, 402, 343

\bibitem[{{Takeda} {et~al.}(2002){Takeda}, {Ohkubo}, \&
  {Sadakane}}]{2002PASJ...54..451T}
{Takeda}, Y., {Ohkubo}, M., \& {Sadakane}, K. 2002, \pasj, 54, 451

\bibitem[{{Takeda} {et~al.}(2005){Takeda}, {Ohkubo}, {Sato}, {Kambe}, \&
  {Sadakane}}]{2005PASJ...57...27T}
{Takeda}, Y., {Ohkubo}, M., {Sato}, B., {Kambe}, E., \& {Sadakane}, K. 2005,
  \pasj, 57, 27

\bibitem[{{Takeda} {et~al.}(2008){Takeda}, {Sato}, \&
  {Murata}}]{2008PASJ...60..781T}
{Takeda}, Y., {Sato}, B., \& {Murata}, D. 2008, \pasj, 60, 781

\bibitem[{{Teske} {et~al.}(2014){Teske}, {Cunha}, {Smith}, {Schuler}, \&
  {Griffith}}]{2014ApJ...788...39T}
{Teske}, J.~K., {Cunha}, K., {Smith}, V.~V., {Schuler}, S.~C., \& {Griffith},
  C.~A. 2014, \apj, 788, 39

\bibitem[{{Thiabaud} {et~al.}(2015){Thiabaud}, {Marboeuf}, {Alibert}, {Leya},
  \& {Mezger}}]{2015A&A...580A..30T}
{Thiabaud}, A., {Marboeuf}, U., {Alibert}, Y., {Leya}, I., \& {Mezger}, K.
  2015, \aap, 580, A30

\bibitem[{{Tull} {et~al.}(1995){Tull}, {MacQueen}, {Sneden}, \&
  {Lambert}}]{1995PASP..107..251T}
{Tull}, R.~G., {MacQueen}, P.~J., {Sneden}, C., \& {Lambert}, D.~L. 1995,
  \pasp, 107, 251

\bibitem[{{Udry} {et~al.}(2002){Udry}, {Mayor}, {Naef}, {Pepe}, {Queloz},
  {Santos}, \& {Burnet}}]{2002A&A...390..267U}
{Udry}, S., {Mayor}, M., {Naef}, D., {et~al.} 2002, \aap, 390, 267

\bibitem[{{Udry} \& {Santos}(2007)}]{2007ARA&A..45..397U}
{Udry}, S. \& {Santos}, N.~C. 2007, \araa, 45, 397

\bibitem[{{Valenti} \& {Fischer}(2008)}]{2008PhST..130a4003V}
{Valenti}, J.~A. \& {Fischer}, D.~A. 2008, Physica Scripta Volume T, 130,
  014003

\bibitem[{{van Leeuwen}(2007)}]{2007A&A...474..653V}
{van Leeuwen}, F. 2007, \aap, 474, 653

\bibitem[{{Ward}(1997)}]{1997Icar..126..261W}
{Ward}, W.~R. 1997, \icarus, 126, 261

\bibitem[{{Wright} {et~al.}(2011){Wright}, {Fakhouri}, {Marcy}, {Han}, {Feng},
  {Johnson}, {Howard}, {Fischer}, {Valenti}, {Anderson}, \&
  {Piskunov}}]{2011PASP..123..412W}
{Wright}, J.~T., {Fakhouri}, O., {Marcy}, G.~W., {et~al.} 2011, \pasp, 123, 412

\bibitem[{{Wright} {et~al.}(2012){Wright}, {Marcy}, {Howard}, {Johnson},
  {Morton}, \& {Fischer}}]{2012ApJ...753..160W}
{Wright}, J.~T., {Marcy}, G.~W., {Howard}, A.~W., {et~al.} 2012, \apj, 753, 160

\bibitem[{{Wright} {et~al.}(2009){Wright}, {Upadhyay}, {Marcy}, {Fischer},
  {Ford}, \& {Johnson}}]{2009ApJ...693.1084W}
{Wright}, J.~T., {Upadhyay}, S., {Marcy}, G.~W., {et~al.} 2009, \apj, 693, 1084

\bibitem[{{Wu} \& {Lithwick}(2011)}]{2011ApJ...735..109W}
{Wu}, Y. \& {Lithwick}, Y. 2011, \apj, 735, 109

\bibitem[{{Wu} \& {Murray}(2003)}]{2003ApJ...589..605W}
{Wu}, Y. \& {Murray}, N. 2003, \apj, 589, 605

\end{thebibliography}

%<<<<<<<<<<<<<<<<<<<<<<<<<<<<<<<<<<<<<<<<<<<<<<<<<<<<<<<<<<<<<<<<<<<<
%                      Apendix
%>>>>>>>>>>>>>>>>>>>>>>>>>>>>>>>>>>>>>>>>>>>>>>>>>>>>>>>>>>>>>>>>>>>>
\begin{appendix}
\section{Additional tables}

% +++ Table with stellar paramters
%\input{tablas/hot_cool_stars_tabla_parameters}

% ---------------------------------------------------------------------------
% Articulo abundancias quimicas en estrellas con hot/cool Jupiters
% ultimo cambio: Junio 02, 2017
% --------------------------------------------------------------------------- 
% Tabla 1: Parametros basicos
% ---------------------------------------------------------------------------

%\onllongtab{
\longtab[1]{
\begin{landscape}
\begin{longtable}{llccccccccc} 
\caption{
 Spectroscopic parameters with uncertainties
 for the stars measured in this work. %Columns 7 and 9
  }\label{parameters_table_full}\\
\hline
 HIP &   Sample$^{\star}$  &  T$_{\rm eff}$ &  $\log g$            & $\xi_{t}$           & [Fe/H] & Sp.$^{\dag}$   &  Age  & M$_{\star}$   & R$_{\star}$ &  Kin$^{\ddag}$   \\
     &                     &       (K)      & (${\rm cm s^{-2}}$)  & (${\rm km s^{-1}}$) & (dex)    &                & (Gyr) & (M$_{\odot}$) & (R$_{\odot}$)  & \\
\hline
\endfirsthead
\caption{Continued.} \\
\hline
HIP &   Sample$^{\star}$  &  T$_{\rm eff}$ &  $\log g$            & $\xi_{t}$           & [Fe/H] & Sp.$^{\dag}$   &  Age  & M$_{\star}$   & R$_{\star}$ &  Kin$^{\ddag}$    \\
     &                     &       (K)      & (${\rm cm s^{-2}}$)  & (${\rm km s^{-1}}$) & (dex)    &                & (Gyr) & (M$_{\odot}$
) & (R$_{\odot}$) & \\
\hline
\endhead
\hline
\endfoot
\hline
\endlastfoot
% ---------------------------
% data comes here
% ---------------------------
\hline\noalign{\smallskip}
522     &       c       m       &       6282    $\pm$   30      &       4.16    $\pm$   0.05    &       1.54    $\pm$   0.16    &       0.07    $\pm$   0.02    &       1       &       2.56    $\pm$   0.23    &       1.25    $\pm$   0.01    &       1.38    $\pm$   0.02    &       D       \\
1292    &       c               &       5566    $\pm$   25      &       4.69    $\pm$   0.05    &       1.27    $\pm$   0.12    &       0.11    $\pm$   0.02    &       1       &       0.38    $\pm$   0.27    &       0.98    $\pm$   0.01    &       0.87    $\pm$   $<$0.01 &       D       \\
1931    &       c               &       5894    $\pm$   20      &       4.37    $\pm$   0.05    &       1.12    $\pm$   0.11    &       0.29    $\pm$   0.02    &       1       &       3.46    $\pm$   0.94    &       1.20    $\pm$   0.04    &       1.34    $\pm$   0.14    &       D       \\
2350    &       h               &       5231    $\pm$   18      &       4.64    $\pm$   0.05    &       0.77    $\pm$   0.13    &       0.20    $\pm$   0.02    &       2       &       2.64    $\pm$   2.66    &       0.92    $\pm$   0.02    &       0.84    $\pm$   0.02    &       D       \\
7513    &       h       m       &       6132    $\pm$   25      &       4.07    $\pm$   0.05    &       1.41    $\pm$   0.11    &       0.06    $\pm$   0.02    &       4       &       3.66    $\pm$   0.47    &       1.26    $\pm$   0.04    &       1.60    $\pm$   0.03    &       D       \\
8770    &       c       m       &       6057    $\pm$   20      &       4.30    $\pm$   0.04    &       1.32    $\pm$   0.10    &       0.25    $\pm$   0.02    &       1       &       2.64    $\pm$   0.35    &       1.22    $\pm$   0.01    &       1.30    $\pm$   0.04    &       D       \\
\end{longtable}
\tablefoot{
 $^{\star}${\bf c}: cool-PH, {\bf h}: hot-PH, {\bf m}: multiplanet system, {\bf m (bd)}: brown dwarf companion.\\  
 $^{\dag}$Spectrograph: {\bf(1)} ESO/FEROS; {\bf(2)} ESO/HARPS; {\bf(3)} SOPHIE; %%% {\bf(2)} ESO/HARPS;
 {\bf(4)} McDonald/2dCoud\'e \\
$^{\ddag}$ {\bf D}: Thin disc, {\bf TD}: Thick disc, {\bf TR}: Transition.
}%
\end{landscape}
}
%%}

% +++ Table with abundances
%\input{tablas/hotcool_table_abundancias}
% ---------------------------------------------------------------------------
% Articulo abundancias quimicas en estrellas con hot/cool planets
% Tabla 2: Abundancias // version 02 de Junio de 2017
% ---------------------------------------------------------------------------

%\onllongtab{
\longtab[2]{
\begin{landscape}
\begin{longtable}{lrrrrrrrrrrrrrrrrrrrr}
\caption{\label{abundance_table_full} Derived abundances [X/H]}\\
\hline\hline
 HIP    &          C~{\sc i}  & O~{\sc i}   & Na~{\sc i} & Mg~{\sc i} & Al~{\sc i}  &
                   Si~{\sc i} & S~{\sc i}   & Ca~{\sc i} & Sc~{\sc i} & Sc~{\sc ii} &
                   Ti~{\sc i} & Ti~{\sc ii} & V~{\sc i}  & Cr~{\sc i} & Cr~{\sc ii} &
                   Mn~{\sc i} & Co~{\sc i}  & Ni~{\sc i} & Cu~{\sc i} & Zn~{\sc i} \\
\hline
\endfirsthead
\caption{continued.}\\
\hline\hline
 HIP    &          C~{\sc i}  & O~{\sc i}   & Na~{\sc i} & Mg~{\sc i} & Al~{\sc i}  &
                   Si~{\sc i} & S~{\sc i}   & Ca~{\sc i} & Sc~{\sc i} & Sc~{\sc ii} &
                   Ti~{\sc i} & Ti~{\sc ii} & V~{\sc i}  & Cr~{\sc i} & Cr~{\sc ii} &
                   Mn~{\sc i} & Co~{\sc i}  & Ni~{\sc i} & Cu~{\sc i} & Zn~{\sc i} \\
\hline
\endhead
\hline
\endfoot
\hline\noalign{\smallskip}
522 &            0.00 &    -0.06 &   0.16 &    0.01 &    0.08 &    0.10 &    0.07 &    0.06 &    0.13 &    0.07 &    0.01 &    -0.01 &   0.09 &    0.06 &    0.05 &    0.10 &    -0.02 &   0.03 &    -0.29 &   -0.07 \\ 
          &  0.06 &       0.05 &       0.08 &       0.11 &       0.07 &       0.01 &       0.34 &       0.05 &       0.05 &       0.07 &       0.03 &       0.06 &       0.08 &       0.04 &       0.06 &       0.05 &       0.04 &       0.02 &       0.04 &       0.09 \\     
1292 &           0.25 &    0.07 &    0.04 &    0.06 &    0.09 &    0.10 &    0.14 &    0.08 &    0.01 &    0.03 &    0.07 &    0.05 &    0.16 &    0.10 &    0.14 &    0.25 &    0.06 &    0.05 &    0.07 &    0.16 \\  
          &  0.14 &       0.05 &       0.07 &       0.03 &       0.04 &       0.02 &       0.10 &       0.05 &       0.07 &       0.05 &       0.05 &       0.05 &       0.04 &       0.04 &       0.07 &       0.05 &       0.05 &       0.02 &       0.04 &       0.17 \\     
1931 &           0.27 &    0.20 &    0.30 &    0.29 &    0.32 &    0.32 &    0.30 &    0.27 &    0.25 &    0.26 &    0.24 &    0.28 &    0.28 &    0.26 &    0.44 &    0.49 &    0.21 &    0.31 &    0.25 &    0.33 \\  
          &  0.17 &       0.03 &       0.06 &       0.03 &       0.01 &       0.01 &       0.12 &       0.04 &       0.05 &       0.05 &       0.03 &       0.05 &       0.02 &       0.03 &       0.13 &       0.04 &       0.04 &       0.02 &       0.11 &       0.16 \\     
2350 &                  &         &  0.24 &    0.19 &    0.22 &    0.21 &    0.29 &    0.06 &    0.24 &    0.19 &    0.22 &    0.18 &    0.30 &    0.21 &    0.18 &    0.39 &    0.24 &    0.21 &           &  0.17 \\  
          &          &            &    0.04 &       0.05 &       0.02 &       0.02 &       0.09 &       0.07 &       0.11 &       0.05 &       0.06 &       0.05 &       0.05 &       0.04 &       0.09 &       0.06 &       0.05 &       0.03 &               &    0.06 \\
\end{longtable}
\end{landscape}
}
%%\end{longtab} 
%}

\end{appendix}
%____________________________________________________________________
%                         FIN
%____________________________________________________________________
\end{document}